\documentclass[format=acmsmall,nonacm, screen]{acmart}
\usepackage{datetime}
\usepackage{graphicx}
\usepackage[linesnumbered,ruled]{algorithm2e}
\AtBeginDocument{%
  }
\usepackage{subcaption}
\usepackage{xspace}
\newcommand{\name}{SAM-GT\xspace}

\setcopyright{acmlicensed}
\copyrightyear{2025}
\acmYear{2025}
\acmDOI{XXXXXXX.XXXXXXX}

\acmJournal{TCPS}
\acmVolume{XX}
\acmNumber{X}
\acmArticle{XXX}
\acmMonth{X}

\acmSubmissionID{TCPS-2024-0063.R1}

\begin{document}

\title{A Scalable Automatic Model Generation Tool for Cyber-Physical Network Topologies and Data Flows for Large-Scale Synthetic Power Grid Models}
\date{\today}

\author{Samantha Israel}
\email{samantha.israel@tamu.edu}
\orcid{0009-0006-7794-4140}
\author{Sanjana Kunkolienkar}
\email{sanjanakunkolienkar@tamu.edu}
\orcid{0009-0001-4539-6075}
\author{Ana Goulart}
\email{goulart@tamu.edu}
\author{Katherine Davis}
\email{katedavis@tamu.edu}
\author{Thomas Overbye}
\email{overbye@tamu.edu}
\affiliation{%
  \institution{Texas A\&M University}
  \city{College Station}
  \state{Texas}
  \country{USA}
}

\renewcommand{\shortauthors}{Israel et al.}

\begin{abstract}
\textit{
Power grids and their cyber infrastructure are classified as Critical Energy Infrastructure/Information (CEII) and are not publicly accessible. While realistic synthetic test cases for power systems have been developed in recent years, they often lack corresponding cyber network models. This work extends synthetic grid models by incorporating cyber-physical representations. To address the growing need for realistic and scalable models that integrate both cyber and physical layers in electric power systems, this paper presents the Scalable Automatic Model Generation Tool (\name). This tool enables the creation of large-scale cyber-physical topologies for power system models. The resulting cyber-physical network models include power system switches, routers, and firewalls while accounting for data flows and industrial communication protocols. Case studies demonstrate the tool's application to synthetic grid models of 500, 2,000, and 10,000 buses, considering three distinct network topologies. Results from these case studies include network metrics on critical nodes, hops, and generation times, showcasing \name’s effectiveness, adaptability, and scalability.}
\end{abstract}

\ccsdesc[500]{Networks~Topology analysis and generation}
\ccsdesc[500]{Networks~Star networks}
\ccsdesc[500]{Networks~Hybrid networks}
\ccsdesc[500]{Networks~Cyber-physical networks}
\ccsdesc[500]{Computer systems organization~Embedded and cyber-physical systems}

\keywords{Electric grid test cases, cyber-physical models, star, radial, statistics-based topology, topology generation, network modeling, cyber-physical model.}

\received[revised]{8 March 2025}
\maketitle
\section{Introduction}
\label{Section:Introduction}
Electric grid models have been used for years to study how power systems function \cite{Spencer1925}. Models or test cases enable researchers to test new algorithms on power grid networks without the risk of interrupting the electricity supply to customers. The information is not publicly available since power grid data is considered critical infrastructure \cite{FERCCEII, EUDirective2008}. Thus, fictitious electric grid models of varying sizes, also known as synthetic models, have been used for research.

Over the past few decades, communication systems have been widely integrated into power systems, transforming them into more intelligent and sophisticated networks. The stable operation of power grids depends on various factors, including the interdependencies between subsystems, system parameters, and control mechanisms—many of which are influenced by the cybersecurity of communication and software systems.

The synergy between cyber and power networks is undeniable; they represent two interconnected layers of the same system. Incorporating communication models, particularly for cybersecurity applications, is crucial for improving our understanding of power grid behavior and informing the development of new planning strategies. This is especially important in the context of high-impact, low-frequency events, such as communication-related failures and cyber threats. 

Cyber-physical models play a critical role in advancing cybersecurity research by enabling researchers to analyze communication nodes, protocols, and adversarial tactics. These models strengthen the resilience and robustness of modern power grids against evolving threats.

Enhancing cyber-physical resiliency benefits a wide range of stakeholders, including utilities, Transmission Service Providers (TSPs), Independent System Operators (ISOs), and consumers. The development of synthetic models, or test cases, provides a strong foundation for advancing practical algorithms that can, in turn, strengthen the cybersecurity of real-world power grid systems.

By building synthetic cyber-physical models, researchers can leverage these datasets to develop and test new applications. These models also serve educational purposes, offering an intuitive way to explore various cyber-physical scenarios. Additionally, they enable visualization techniques that enhance understanding of cyber-physical interactions and system behavior under different adversarial actions.

The availability of test cases with realistic cyber-physical topology is limited. Some cyber-physical models are restricted to distribution systems or smaller IEEE test cases \cite{cyberdist1,cyberdistCosim}. A power distribution cyber-physical test case presented in \cite{CyberDist} considers the correlation between cyber and physical systems by covering a range of power generation and load variations. The 8-substation model \cite{8SubWeaver} is an early example of a cyber-physical test case that models an electrical transmission system, following the Western System Coordinating Council (WSCC) 9-bus model. 

A large-scale cyber-physical model was proposed in \cite{Texas2000}, based on the Texas 2000-bus system. More recently, a statistics-based cyber-physical topology for the European power grid was introduced in \cite{europe_topology}. This model addresses the scalability challenges of representing large cyber-physical networks by incorporating three distinct layers: the power system layer, the communication layer (which includes physical devices such as fiber optic switches), and the logical communication layer (which accounts for network routers). The tool also introduces redundant communication nodes within the communication layer to enhance reliability. However, this model does not capture data flows or the communication protocols used in utilities' Supervisory Control and Data Acquisition (SCADA) systems, and it supports only a single network topology.

In this paper, the creation of a cyber-physical model of a power grid is automated. The goal of this approach is to take any power system network model as input and generate a realistic communication topology that overlays the physical power system layer. This topology is designed to be valid alongside the power system model for joint simulation and/or emulation. It can also be customized based on the specific use cases being studied.

To demonstrate this, cyber-physical topologies are generated and validated using metrics for three synthetic power grid models, with bus sizes ranging from 500 to 10,000. These cyber-physical models include relays, substations, Utility Control Centers (UCC), and Balancing Authorities (BA), developed based on the approach in \cite{Texas2000}. Creating such a versatile framework holds substantial promise for advancing the field of power grid modeling and enhancing cyber-physical system integration.
Specifically, the creation of these cyber-physical synthetic grid models is of crucial importance for the development and validation of novel applications in power system critical infrastructure defense including cyber-physical situational awareness techniques and intrusion response engines.

In summary, the main contributions of this paper are as follows:
\begin{itemize}
    \item The work presents \name, a scalable turn-key approach for generating large-scale realistic cyber-physical models. The model generation tool produces a large-scale cyber-physical power system model, with a detailed cyber layer which is overlayed and connected with the large-scale electric power system model. 
    \item \name creates object-oriented data structures for representing the system within and between different power system levels (e.g., substations, utilities, and balancing authorities), where the communication layer models real-world data flows and operations.
    \item \name models the data flows and operations using different communication protocols that would be found at different levels and in different geographical/ownership regions of a power system.
    \item \name is scalable and validated with three synthetic power networks ranging from 500 to 10,000 buses. 
    \item The models are created by \name include high-fidelity representation of the following: 
    \begin{itemize}
        \item Communication protocols; 
        \item Cyber devices such as firewalls and forwarding rules, routers and their internet protocol (IP) configurations, and Ethernet switches;
        \item Special types of hosts found in utilities such as computers that host Human Machine Interface (HMI) or SCADA master or database applications; 
        \item Cyber-physical devices such as relays and relay controllers depending on the number of buses in a substation as configured by the user.
    \end{itemize}
    \item The process of creating the models is automated and configurable. This allows users to generate and test different large-scale realistic cyber-physical power system models under different network topologies, such as star, radial, and a statistics-based topology, and configure the number of control centers and balancing authorities.  It also allows users to generate and test impacts of varying model features including different network topologies and substation layouts on system performance metrics.
\end{itemize}

This paper is organized as follows. Section ~\ref{Section:Current Cases and Models} presents an overview of electric and cyber-physical synthetic models. Section~\ref{Section:Communication Model} describes the communication model. Section~\ref{Section:Algorithm} presents the process of creating the cyber-physical network. Section~\ref{Section:Results} compares statistics of the 500, 2,000 and 10,000-bus models, and discusses our findings and related work. This paper is concluded in Section~\ref{Section:Conclusions}.

\section{Existing Test Cases and Models}
\label{Section:Current Cases and Models}
This section is divided into two subsections: the first provides an overview of synthetic electric grid test cases, while the second reviews cyber-physical test cases.

\subsection{Electric Grid Models and Test Cases}

Protecting power infrastructure is a priority for many countries, as they view their power grid networks as essential to national growth and security. Consequently, the components and architecture of an active power grid are often kept confidential. Test cases have historically played a role in evaluating new algorithms in power system research. Some of these test cases are available in the IEEE format~\cite{UWteseIEEE,pinheiro1998probing}. These test cases are often small; they have fewer buses and lack geographical information. Although these test cases have contributed significantly to seminal research, there is a growing need to study electrical networks on a larger scale, better mimicking real-world power grids. Several large-scale test cases have emerged in the past decade\cite{sadeghian2020autosyngrid,SnodgrassTractable2021, chatzos2022data, gegner2016methodology, birchfield2016grid}. Since these cases have fictitious information, they are called synthetic electric grids. Their goal is to mimic the behavior of a real-world power system network, relevant data, such as geographic locations and population density, are extracted from census information to guide the creation of realistic synthetic grids. 

\begin{figure}[]
\begin{minipage}[h]{0.32\linewidth}
\includegraphics[width=\columnwidth]{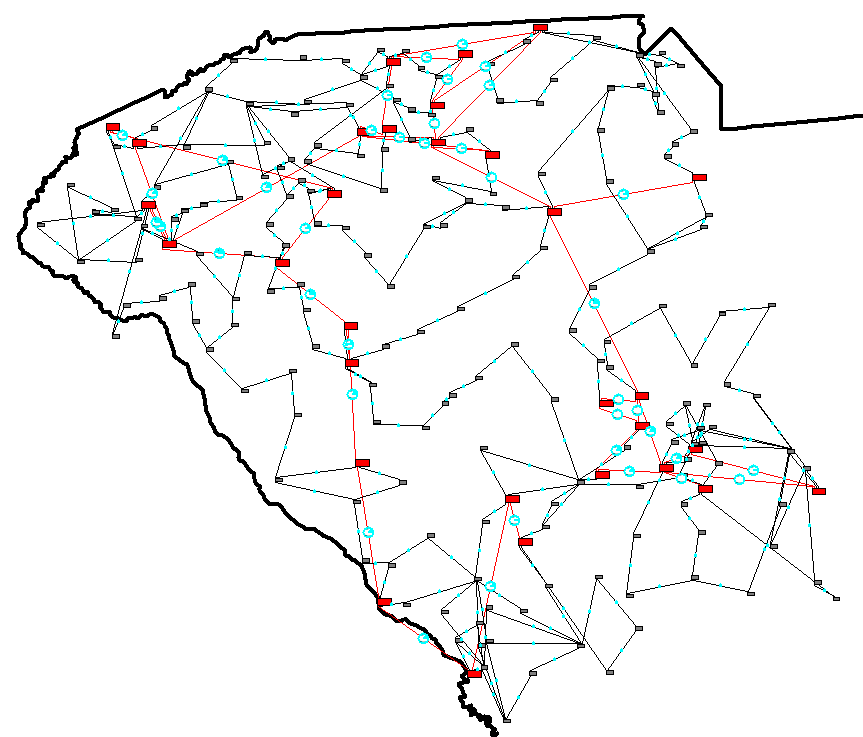}
\end{minipage}
\hfill
\begin{minipage}[h]{0.32\linewidth}
\includegraphics[width=\columnwidth]{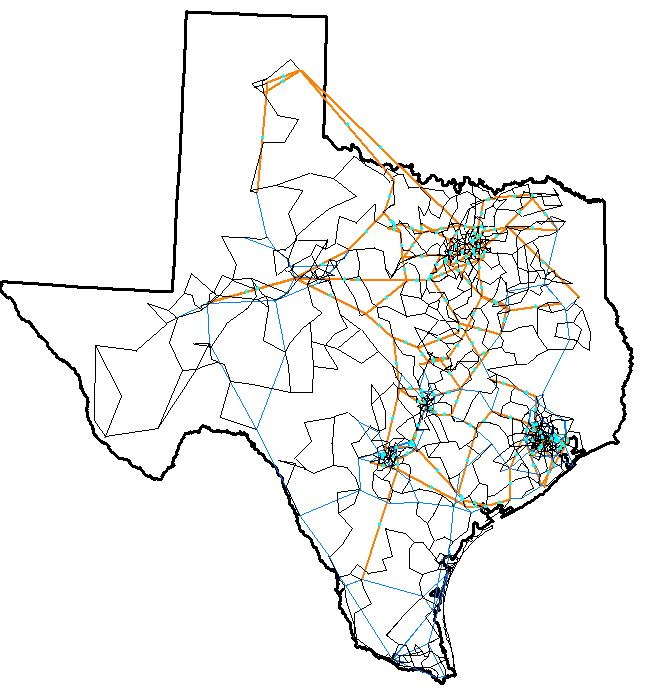}
\end{minipage}
\hfill
\begin{minipage}[h]{0.32\linewidth}
\includegraphics[width=\columnwidth]{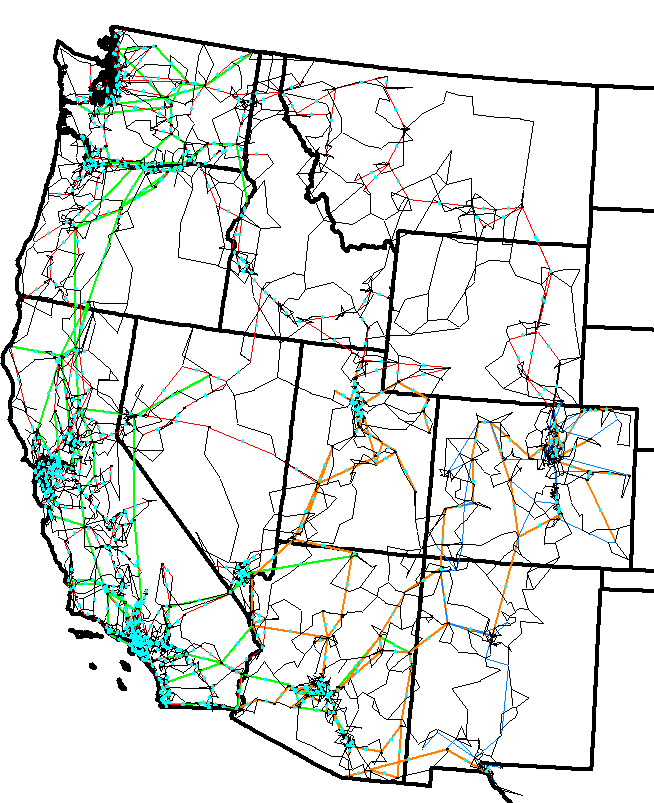}
\end{minipage}
\caption{Large-scale synthetic electric grid models: (left) 500-Bus (footprint of South Carolina); (center) 2,000-Bus (footprint of Texas); (right) 10,000-Bus (footprint of Western U.S.)}
\label{fig:allnetworks}
\end{figure}

The electric grid models contain information about the power network: buses with loads, generators, and transmission lines connecting these buses. The synthetic networks used in this work can be found at \cite{TAMUrepository}. This repository's power system test cases contain a few hundred to 80,000 buses associated with substations containing geographical information. While the algorithm in this paper to build a cyber-physical model can be applied to a power grid synthetic model of any size, the results are shown for three synthetic cases as described below, also seen in Fig.\ref{fig:allnetworks}: 
\begin{itemize}
    \item \textit{the 500-bus case on the footprint of South Carolina}: This case has 500 buses, 90 generators, 206 loads, and 468 transmission lines. It has a total of 208 substations. For this case, the assumption made in this paper is that it has \textbf{4} utilities and \textbf{1} balancing authority.
    \item \textit{the 2,000-bus case on the footprint of Texas}: This case has 2,000 buses, 544 generators, 1,356 loads, and 2,345 transmission lines. It has a total of 1,250 substations. For this case, the assumption made in this paper is that it has \textbf{20} utilities and \textbf{1} balancing authority.
    \item \textit{the 10,000-bus case on the footprint of the Western United States}: This case has 10,000 buses, 2,485 generators, 4,899 loads, and 9,726 transmission lines. It has a total of 4,762 substations. For this case, the assumption made in this paper is that it has \textbf{80} utilities and \textbf{20} balancing authorities.
\end{itemize}

\subsection{Cyber-Physical Models and Test Cases}
Cyber-physical models not only aid in assessing system performance but also help evaluate the impact of cyber threats on power system operations and the implementation of appropriate security measures. Understanding network topologies and their implications for cyber-physical systems is essential to ensuring the integrity and functionality of electrical power grids. As a result, various cyber-physical models for power grid networks have been developed for research.

One common approach is to model the communication network using the electric grid test case as a reference, resulting in similar topologies for both networks. This method was demonstrated in \cite{decentralized}, where a communication model was developed based on the IEEE 118-bus synthetic grid. In that study, the communication network was represented as a graph, with substations as nodes and communication lines as links, to analyze the impact of latency on Phasor Measurement Unit (PMU) messages.

The study compared a centralized topology, where substations connect directly to a control center, with a decentralized topology that includes an intermediate communication hub. The findings indicated that the decentralized topology resulted in lower delays. The algorithm proposed in this paper also generates both centralized and decentralized communication network topologies but follows a different approach for connecting the hubs compared to \cite{decentralized}.

A large-scale cyber-physical model for a power system is presented in \cite{europe_topology}. That study models the power grids of multiple European countries using a decentralized network topology, where substation routers connect to intermediate communication hubs, which in turn link to a utility control center. The model also incorporates network redundancy and identifies the most critical nodes. The study presents various results and comparisons demonstrating the scalability of its proposed cyber-physical power system model generation approach.

Similarly, in the 8-substation model \cite{8SubWeaver}, the cyber-physical system consists of four interconnected networks: electrical, substation control, secondary substation, and control center. The electrical power network includes the physical components that directly interface with Intelligent Electronic Devices (IEDs). The substation control network contains the IEDs, while the secondary substation network provides the cyber-physical link between the electrical power and substation control networks. The control center network encompasses the Energy Management System (EMS) and the Supervisory Control and Data Acquisition (SCADA) system. Each substation network has its own Java-Object Oriented Notation (JSON) file, which defines nodes and links. These JSON files include nodes such as switches, relays, breakers, buses, transformers, and generators.

Using an object-oriented approach and detailed communication nodes, the authors in \cite{Texas2000} model a star topology where all substations within a given region of the Texas grid connect to a Utility Control Center (UCC). The placement of UCCs is determined by identifying a central node through the K-nearest neighbor method, using the geographic coordinates of substations. The model also includes a Regulatory Authority or Balancing Authority (BA) linked to all UCCs, with each site incorporating communication infrastructure such as routers, firewalls, switches, and relays. In \cite{Stats}, statistical metrics were collected from a simplified communication network topology based on a realistic utility communication system presented in \cite{osti_1526728}.

\section{Communication Model Design}
\label{Section:Communication Model}

\begin{figure*}
\centering
    \includegraphics[width=\linewidth]{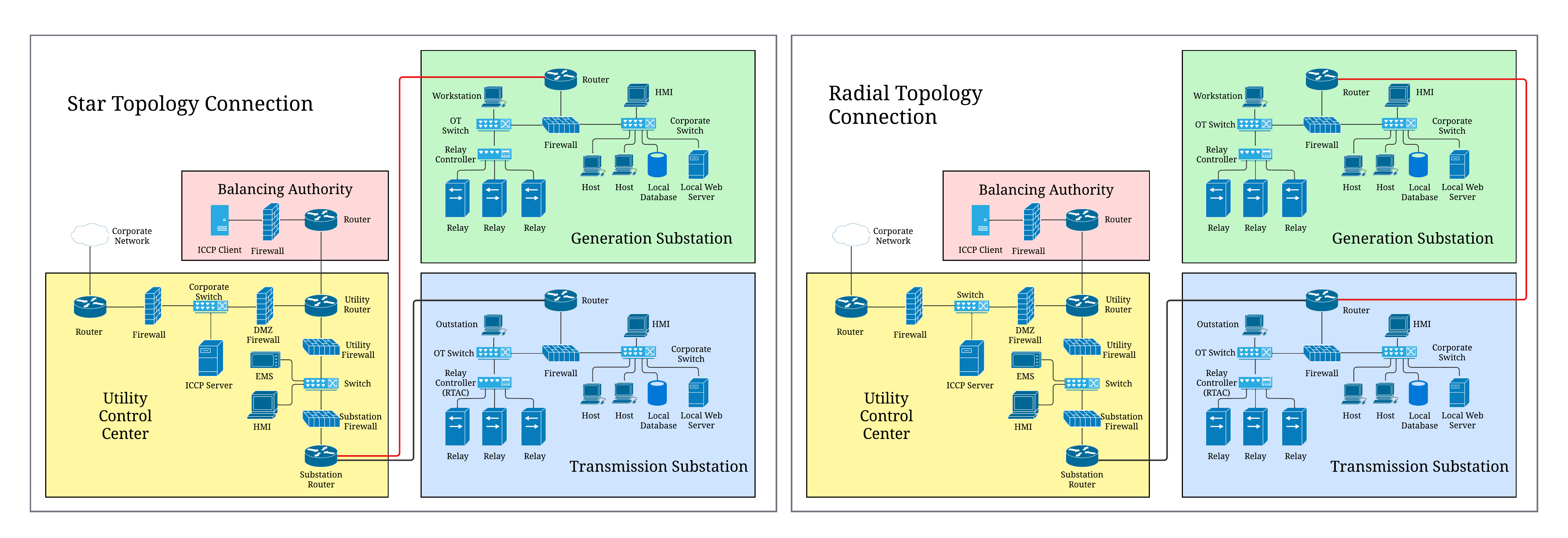}
    \caption{Network topology based on~\cite{Texas2000}.  This shows two different topologies, star, and radial, and how the different substations would connect to the UCC.}
    \label{fig:star&radial}
\end{figure*}

Fig.~\ref{fig:star&radial} delineates the architectural framework of the proposed communication model, illustrating the network configurations for a UCC, a BA, and two distinct substation types: generation and transmission. Fig.~\ref{fig:star&radial} demonstrates how these entities are interconnected using two prevalent Wide Area Network (WAN) topologies: star and radial. Each substation is directly connected to the UCC in the star topology, facilitating centralized control. Conversely, the radial topology employs a hierarchical structure where generation substations link to transmission substations, which connect to the UCC. The nuances and implications of these topologies and a third topology are further elaborated in the subsequent subsection.

\subsection{Modelling Network Topologies}

\begin{figure*}
\centering
    \includegraphics[scale=0.155]{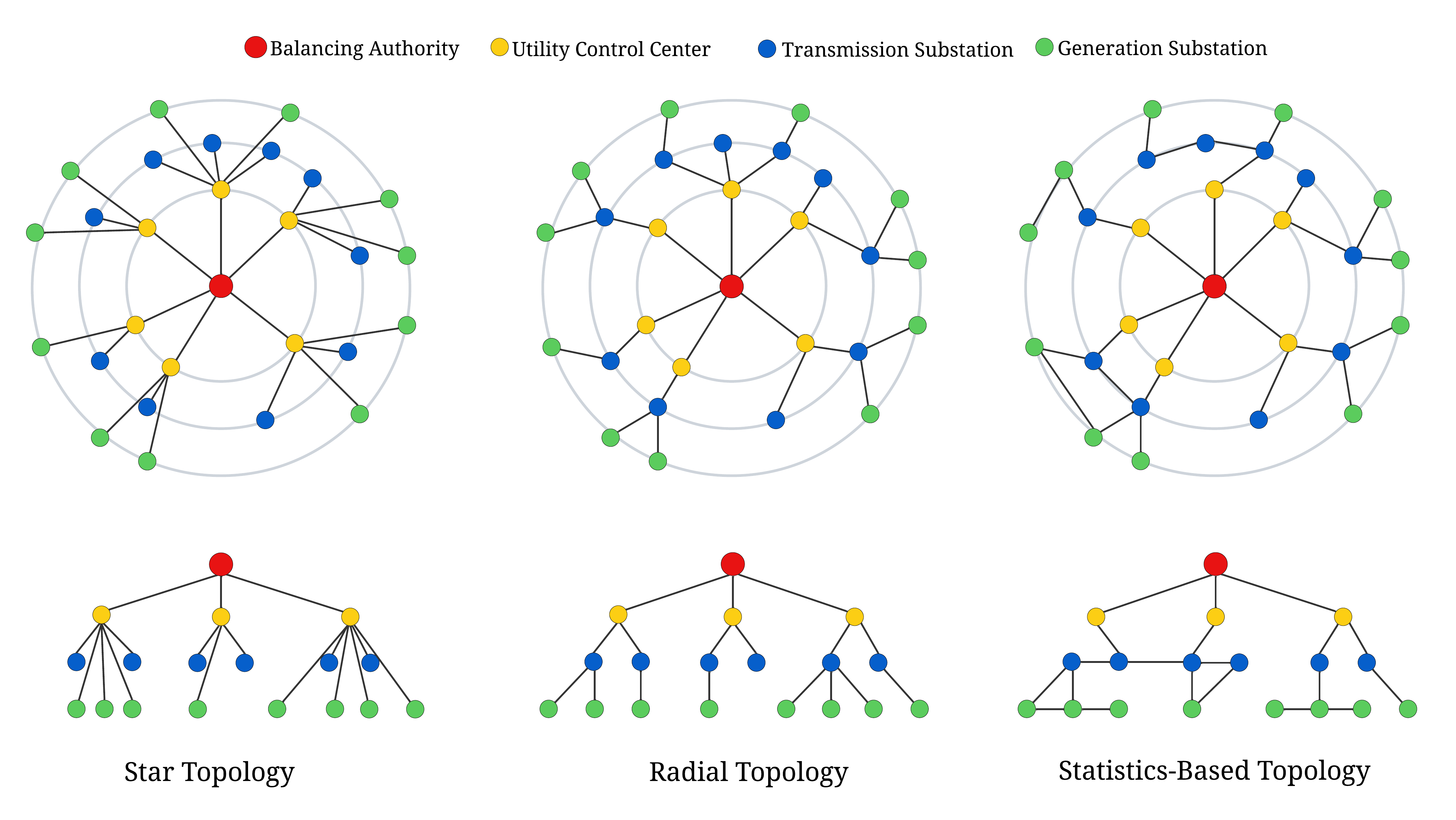}
    \caption{Comparison between star, radial, and statistics-based topologies illustrates how the substations are connected to the UCC.}
    \label{fig:topologyExplanation}
\end{figure*}

The network topology in the study of Cyber-Physical Power Systems (CPPS) may refer to the topology of the physical system, i.e., the power grid with its substations, buses, and transmission lines. It may also refer to the communication network and how communication links connect routers, firewalls, switches, and hosts. In this paper, the network topology refers to the communication network topology. The proposed model generation focuses on star, radial, and statistical topologies. The difference among these topologies is how the substations are connected to the UCC. These network topologies are illustrated in Fig.~\ref{fig:topologyExplanation}, where each node matches the color of the communication network shown in Fig.~\ref{fig:star&radial}. The choice of topology depends on factors such as the devices to be connected, the desired reliability of the network, and the cost of cabling\cite{networkBasics}. Each topology offers different advantages and is employed in specific contexts based on scalability, fault tolerance, and ease of maintenance\cite{Metamodel}. Following is an overview of each topology: 

\begin{itemize}
\item Star topologies have all nodes connected directly to a central node, such as a UCC. All data passes through the central node. If it fails, communication between all substations is disrupted.

\item Radial topologies organize nodes in a hierarchical structure, where a central node connects to one or more secondary nodes, which in turn link to additional nodes. Since generation and transmission substations serve different functions, their connections to utilities can vary, making this hierarchical structure a suitable approach. However, a key drawback of radial topologies is their susceptibility to single points of failure\cite{RadialLayout}.

\item Statistics-based: The communication links can be modeled after the transmission power lines between substations, but the communication network does not exactly follow the power network in the real world. Hence, this paper also considers that substations can be connected in a mesh/hybrid topology that we call statistical topology connection, based on the network metrics derived in~\cite{Stats}. 

\end{itemize}

\subsection{Modeling Utility Control Centers (UCCs) and Balancing Authorities (BAs)}

In synthetic power networks, geographical data is typically associated with substations, whereas information about UCCs or BAs is often considered administrative and not crucial for power flow analyses. Nevertheless, the critical role of communication networks connecting substations, UCCs, and BAs necessitates the inclusion of UCC and BA locations within the analytical model. The methodology introduced in this paper initially establishes the positions of the UCC and BA. Subsequently, it integrates the communication model details and allocates data flows to each UCC and BA.

As the goal is to construct a realistic yet fictional network topology, the location of a UCC is inferred from the latitude and longitude of the substations. This is achieved through $k$-means clustering, which calculates the centroids of substations. The value of $k$, representing the number of clusters, is determined by user input. For example, in the 500-bus grid model, the number of UCCs is 4, that is $k$ = 4. The position of the regulatory agency or BA also needs to be determined. A BA's role is to ensure the generated power meets the consumers' load. A BA also controls the power grid's frequency, to keep it stable around the nominal frequency (e.g., 60 Hz in the US). The model uses a user input for the number of BAs. The location of the BA is decided based on the clustering of the UCCs. The BA is placed within the clusters by calculating the centroid of the UCC locations.

The UCCs and BAs contain routers, firewalls, and ethernet switches. Since the UCC oversees its substations, it employs a SCADA system to remotely monitor and control them. The hosts in the UCC include an EMS, a Human Machine Interface (HMI), and a dedicated host to exchange data with the BA. The UCC also has a firewall connected to the corporate network, but this communication network has not been added to our model. The LAN for all UCCs and BAs is shown in Fig. \ref{fig:star&radial}, irrespective of the choice of topology.

\subsection{Modeling Substations}

Information on substations is obtained from the power network data. The location is determined by the latitude and longitude contained in the CSV file. On the communication side, the LAN inside the substations is defined as indicated in Fig. \ref{fig:star&radial}, which is the same for all the substations in the network. They differ in the number of relays and, hence, the connections of relays to the relay controller. Each substation is connected to the UCC's SCADA system or other substations (in the case of radial/statistics-based topology) through a router and a firewall.

The assumption is that each substation has two LANs. In Fig.~\ref{fig:star&radial}, considering the transmission or generation substation, the network on the right of the firewall can be viewed as an administrative network. It includes two hosts, a local database, and a local web server that stores power system data. The network on the left of the firewall is the Operational Technology (OT) network with an outstation host and several relays, depending on the number of buses in the substation. This network also has a relay controller and one outstation host. 

For simplicity, it is assumed that the relay controller combines the functions of a Remote Terminal Unit (RTU) device. The choice of communication link connecting substations to other sites depends on factors like latency, location, security, and cost. Various technologies, such as cellular modems, microwave links, and fiber optics, can be employed. Each link connecting the cyber elements has attributes such as link type, distance, bandwidth, and protocol to represent the diverse communication methods used by utilities.

\subsection{Modeling Data Flows}
Before the network topology is developed and the firewalls are configured, the data flows between the different cyber nodes in the WAN need to be identified. These data flows represent the applications or services running at each node and the data they exchange. The data consists of measurements and commands transmitted between the substation and the UCC and between the UCC and the BA. The data can have many formats, depending on the application-layer communication protocol. These protocols are usually encapsulated in TCP/IP packets. Including the data flows in our model is essential because if a link or node fails due to a physical problem or a cyber attack, the impacted nodes can be observed.

\begin{figure}
    \centering
    \includegraphics[width=\columnwidth]{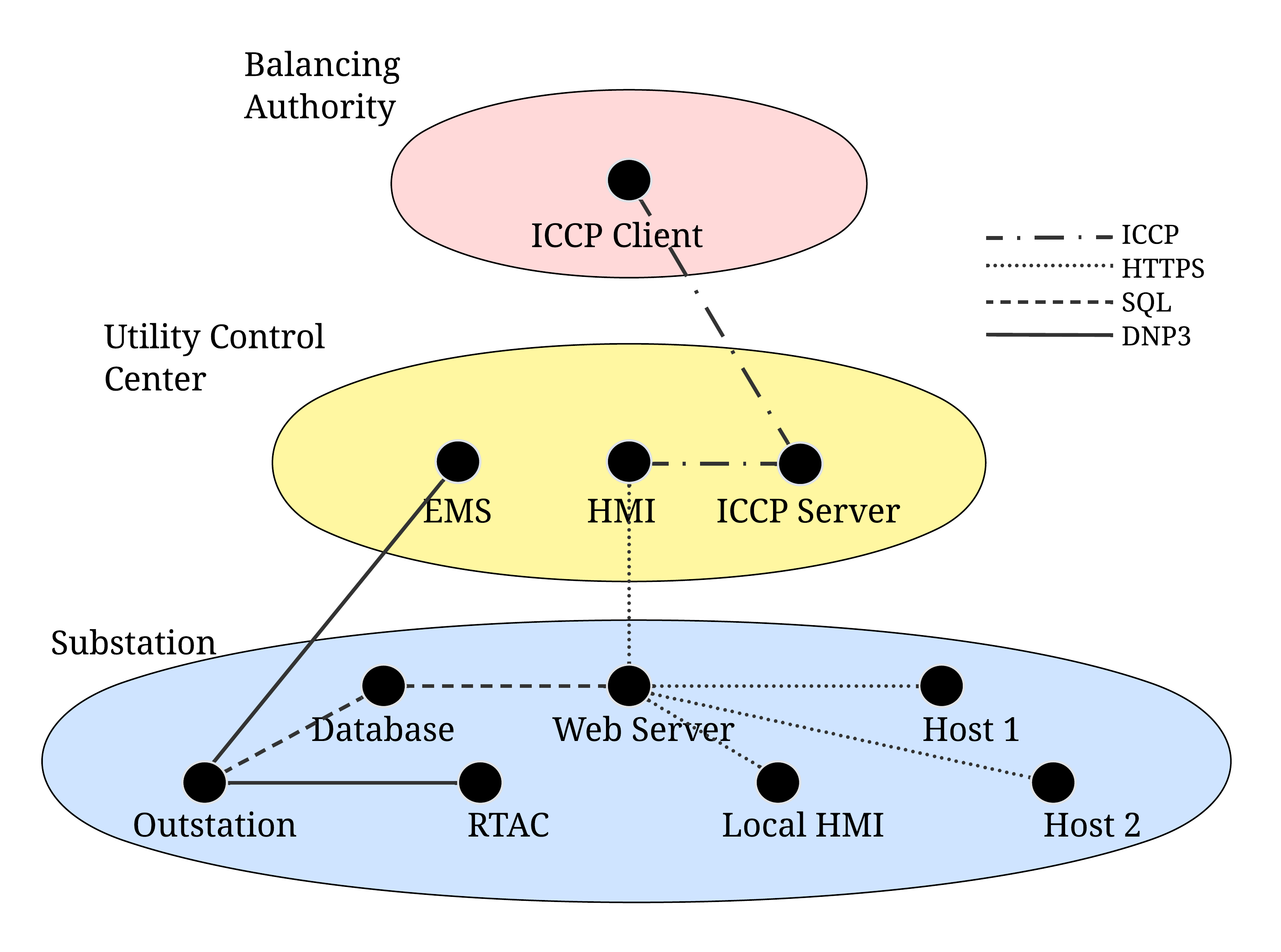}
    \caption{Data flows used between nodes.}
    \label{fig:protocols}
\end{figure}

Fig.~\ref{fig:protocols} illustrates the WAN nodes, e.g., substation, UCC, and BA, and the data flows between them. In each node, specific cyber components communicate with each other based on the network architecture and data flows described in~\cite{TashasThesis}. The cyber components have assigned protocols depending on their function. The current model assumes the following four communication protocols: 

\begin{itemize}
\item Inter-Control Center Communications protocol (ICCP)~\cite{iccp_paper} is used for communication between control centers and a regulatory agency (balancing authority). For example, one of the roles of the regulatory is to communicate with utilities and make sure the power grid's voltage and frequency remain stable. It operates over a wide area network. Balancing authorities can read values and send grid operation commands to utilities. Typically, the utility has an ICCP server configured to send periodic ICCP reports to the balancing authority. 

The 2,000-bus use case, which is a synthetic model of the Texas' power grid, has one balancing authority called Electric Reliability Council of Texas (ERCOT). All market participants in the Texas grid report to ERCOT using ICCP protocol\cite{ERCOT2024}.

\item Distributed Network Protocol 3 (DNP3) is an industrial control system (ICS) protocol designed for SCADA systems. A typical scenario for DNP3 is for the SCADA master to read values and send commands to field devices in the substation. Inside the substation, we assume the communication between the outstation node and the Real-Time Automation Controller (RTAC) also uses DNP3 protocol.

\item Secure Hypertext Transfer Protocol (HTTPS) can be used inside a substation and between substation and the utility control center. A typical scenario would be for a substation to have a local web-based server that can transfer encrypted data to the human-machine interface (HMI) host in the utility control center. Details of this data flow can be found in \cite{TashasThesis}.

\item Structured Query Language (SQL) is a database protocol for uploading or retrieving substation data from the outstation to the local database, which the web server can access. Another scenario is that utilities use SQL databases located for example in a demilitarized zone where utility managers or even customers can monitor the utility's data, such as load or power generation.
\end{itemize}

In the proposed model in this work, firewalls are strategically positioned at substations, UCCs, and BAs to monitor and regulate network traffic. A firewall acts as a barrier that filters both inbound and outbound communications within the network. For instance, a firewall located at a substation might be configured with a rule that specifically allows incoming DNP3 requests from the utility control center. This configuration is illustrated in Fig.~\ref{fig:protocols}, where the model incorporates rules defined by the firewall Access Control Lists (ACLs).

\section{Algorithm}
\label{Section:Algorithm}

\begin{figure}[b]
    \centering
    \includegraphics[width=0.95\columnwidth]{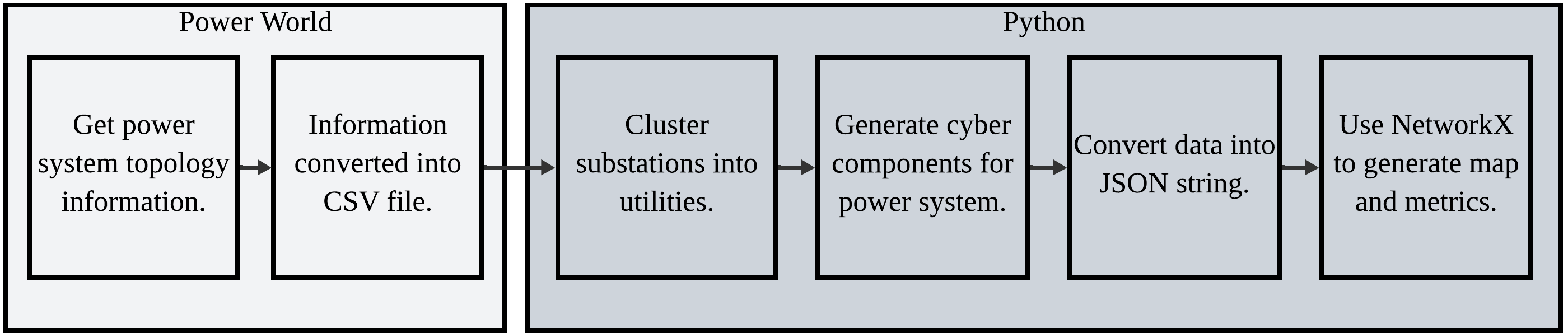}
    \caption{Process for creating the cyber-physical JSON model.} 
    \label{fig:pseudocode}
\end{figure}

This section describes how the proposed tool, \name,  generates the cyber topology given any electric grid model. Fig.~\ref{fig:pseudocode}, provides an overview of the algorithm which is mainly divided into two phases: obtaining power system data (from PowerWorld) and creating the cyber-physical network (using Python). 

\subsection{Power Network Data}
\label{sec:JSONGen}

\begin{table*}
\centering
\caption{Sample rows of CSV file for South Carolina's 500-bus model, for 208 substations.}
\scalebox{0.8}{
\begin{tabular}{c c c c c c} 
\hline
\textbf{Lon.} & \textbf{Lat.} & \textbf{Sub\#} & \textbf{Sub Name} & \textbf{Area Name} & \textbf{Zone}\\
\hline
\hline
-81.09 & 34.37 & 1 & WINNSBORO & SouthCarolina & Columbia \\
\hline
-80.89 & 34.14 & 2 & COLUMBIA 11 & SouthCarolina & Columbia \\
\hline
-81.39 & 35.01 & 3 & SMYRNA & SouthCarolina & York \\
\hline
-80.62 & 33.83 & 4 & EASTOVER 2 & SouthCarolina & Columbia \\
\hline
... & ... & ... & ... & ... & ... \\
\hline
-82.25 & 33.93 & 208 & MC CORMICK & SouthCarolina & Midlands\\
\hline
\end{tabular}
}
\label{csva}
\end{table*}

\begin{table*}
\centering
\caption{Sample rows of CSV file for South Carolina's 500-bus model, a continuation from Table \ref{tab:csva}.}
 \scalebox{0.8}{\begin{tabular}{c c c c c c c c} 
 \hline
 \textbf{Sub\#} & \textbf{...} &\textbf{\#Buses} & \textbf{kV} & \textbf{Gen(MW)} & \textbf{Gen(Mvar)}  & \textbf{Load(MW)} & \textbf{Load(Mvar)}\\ [0.5ex] 
 \hline
\hline

1 & ... & 2 & 138 &  &  & 30.2 & 8.05\\
\hline
2 & ... & 2 & 138 &  &  & 93.62 & 24.97\\
\hline
3 & ... & 2 & 138 &  & & 2.18 & 0.58\\
\hline
4 & ... & 3 & 345 & 231.54  & 36.67 & & \\
\hline

... & ... & ... & ... & ... & ... & ... & ...\\
\hline

208 & ... & 1 & 138 &  &  & 16.27 & 4.34\\
\hline
\hline

\end{tabular}}

\label{tab:csvb}
\end{table*}

\begin{table}[t]
\centering
\caption{500-bus model - Branches information.}
\scalebox{0.8}{\begin{tabular}{c c} 
 \hline
 \textbf{SubNumberFrom} & \textbf{SubNumberTo} \\ [0.5ex] 
 \hline
\hline
1 & 1 \\
\hline
19 & 1 \\
\hline
167 & 1 \\
\hline
2 & 2 \\
\hline
26 & 2 \\
\hline
2 & 195 \\
\hline
\end{tabular}}
\label{tab:csv2}
\end{table}

Using PowerWorld, the substation and branch data files are exported as two Comma Separated Value (CSV) files.
\begin{itemize}
\item Substation data: Tables~\ref{tab:csva} and~\ref{tab:csvb} provide examples of substations' locations, identifiers, and other attributes for the 500-bus case. Substations are classified into generation and transmission. Generation substations list their power outputs under the \emph{Gen MW} and \emph{Gen Mvar} columns, which denote generated megawatts and megavolt-amperes reactive. Substations without generation output are identified as transmission substations. 

\item Branches data: Table~\ref{tab:csv2} shows the branch information of selected substations in the 500-bus case. It lists all transmission lines between two substations. Our tool uses this file to connect transmission and generation substations in the radial topology.
\end{itemize}

While the CSV file also has load information in megawatts, the proposed model does not currently use this information. However, it shows that future distribution substations can also be created to be included in the cyber-physical model.
\begin{figure}[b]
\centering
\includegraphics[scale=0.2]{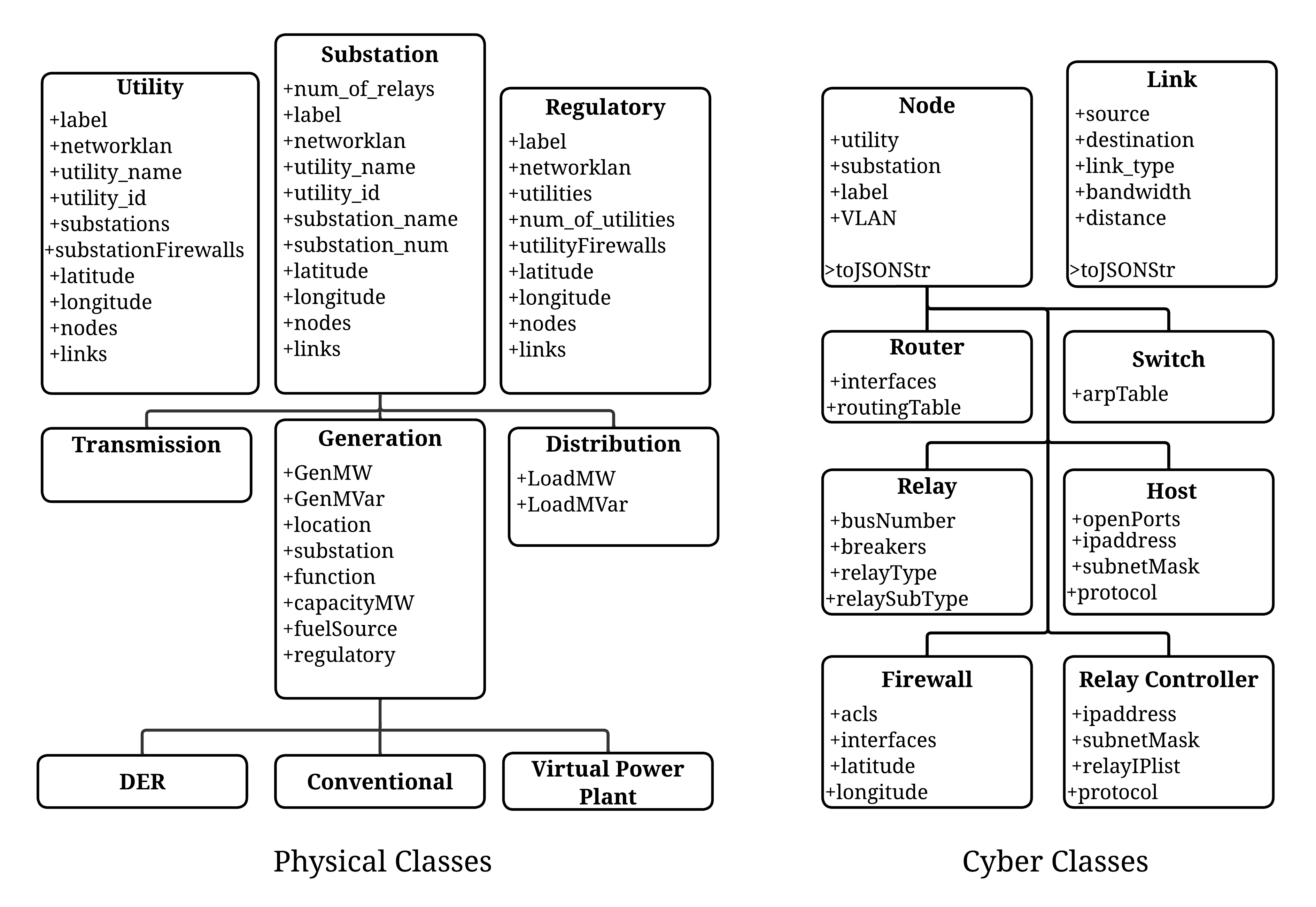}
\caption{Cyber and physical classes.}
\label{fig:cyphys_classes}
\end{figure}
\subsection{Creation of the cyber-network model}
The code starts by reading user inputs from the \emph{settings.ini} file, which has parameters such as the number of UCCs, the number of BAs, the name of the synthetic power system, and the topology (e.g., star, radial, or statistical).

Once these configuration parameters are defined, the process follows the steps in  Fig.~\ref{fig:pseudocode}: cluster substation into utilities, generate cyber components, convert data to JSON, and visualize network maps and metrics using NetworkX. These steps are described in detail next. Algorithms~\ref{alg:generate_substations} and ~\ref{alg:generate_utilities} show the pseudocode to create substations, UCCs, and BAs.

\subsubsection{\textbf{Cluster Substations into Utilities}}
Substations are allocated to UCCs using a $k$-means clustering algorithm, with $N$ representing the number of utilities specified by the user in the \emph{settings.ini} file. The centroids of these clusters determine the placement of UCCs within star and radial topologies. After clustering, labels are assigned to all nodes, including utilities and substations. For example, a substation label looks like this: \emph{Region.Utility\_number.Substation\_name}. This label indicates the substation’s regional and utility affiliation. Regardless of the topology, each substation, UCC, and BA have specific cyber components, as depicted in Fig.~\ref{fig:star&radial}. These cyber components are created using an object-oriented programming approach.

\subsubsection{\textbf{Creating Cyber-Physical Components}}
The cyber-physical model is created using object-oriented programming based on the architecture in Fig.~\ref{fig:cyphys_classes}. It has five classes (data types) including \emph{Substation}, \emph{Utility}, \emph{Regulatory}, \emph{Nodes}, and \emph{Links}. They are organized as \emph{Cyber} Classes and \emph{Physical} Classes.

\emph{Links} and \emph{Nodes} are considered \emph{Cyber} classes. The \emph{Link} class has attributes such as the link's source and destination nodes, the media type of the link, bandwidth in bits per second, and distance (or length of the link). The \emph{Node} class contains attributes such as region, utility, substation, IP address, label, and the Virtual Local Area Network (VLAN) where the node is installed. 
Its sub-classes generate the cyber and cyber-physical nodes inside substations and utilities, such as switch, router, host, firewall, relay, and relay controller. To indicate the types of applications they support, hosts and relay controllers have attributes such as \emph{openPorts} and \emph{protocol}.
\emph{Physical} classes include \emph{Utility}, \emph{Substation} and \emph{Regulatory}, which are higher-level nodes that contain links and nodes. Their relationship to each other can be explained as follows: a \emph{Utility} contains \emph{Substations}, while a \emph{Regulatory} contains \emph{Utilities}. In particular, a \emph{Substation} class has unique attributes such as the number of relays it contains (\emph{relaynum}). 
Furthermore, we can model different types of substations, such as generation, transmission, and distribution. Generation substations can in turn be modeled as conventional generation, distributed energy resources (DER), or virtual power plant. In this paper, we focus on transmission and conventional generation substations.
Being part of the Wide Area Network, utility, substation and regulatory objects need some type of address. For this reason, they are configured as Internet Protocol (IP) networks, with an IP version 4 address and subnetwork mask, as defined by the attribute \emph{networklan}. This IP configuration allows us to segment a utility into Operational Technology (OT) network, routing network, and corporate network. These three types of networks are collectively known as the SuperNet network.
\name's ability to create cyber and physical objects allows us to model nodes and links with specialized functions. Each function is relevant to the type of network the node belongs to; for example, an OT network would have specialized nodes such as relays and relay controllers. Once all the objects are created, the algorithm outputs all the elements with their attributes into a JSON file.
\subsubsection{\textbf{Visualizing Network and Validating Metrics}}
After SAM-GT generates the JSON files, it creates graphs to enhance the visualization of the cyber-physical network. The graphs are plotted using NetworkX software tool, with JSON files as inputs. It generates two graphs: an external view that represents the wide area network, and an internal view that represents the local area networks inside substations, control center, and balancing authority. The results of these WAN and LAN views are presented in Sections~\ref{wan_networkx} and \ref{lan_networkx}.

\begin{algorithm}[H]
\raggedright
\caption{Create Substation Objects and Assign Utilities}
\DontPrintSemicolon
\label{alg:generate_substations}
\SetKwProg{generatesubstations}{Function \emph{create\_substations}}{}{end}
\generatesubstations{$(Substations\_list)$}{
     \For{substation $sub$ in $Substations\_list$}{
         Create an instance of Substation class $sub$ (classify as trans. or gen. station).\;
         Populate instance $sub$ with node objects of cyber classes\;
         [shown in Fig. \ref{fig:cyphys_classes}]\;
         $\textit{nodes} \gets$ [`cyberphysicalNodes', `cyberNodes'].\;
         $\textit{links} \gets$ [Based on the topologies in Fig. \ref{fig:star&radial}]\;
         \textbf{\textit{populate\_cyber\_information()}}\;
         Output $sub$ data into JSON files.\;
         }
    Group substation instances into $N$ clusters (or utilities) using $k$-means clustering based on the latitude and longitude of substations. Assign centroids to utilities. Consider $N$ from the $settings.ini$ file.}
\end{algorithm}

\section{Results and Validation}
\label{Section:Results}

\subsection{Metrics for Analysis} 
\begin{figure}[h]
\centering
\includegraphics[width=\columnwidth]{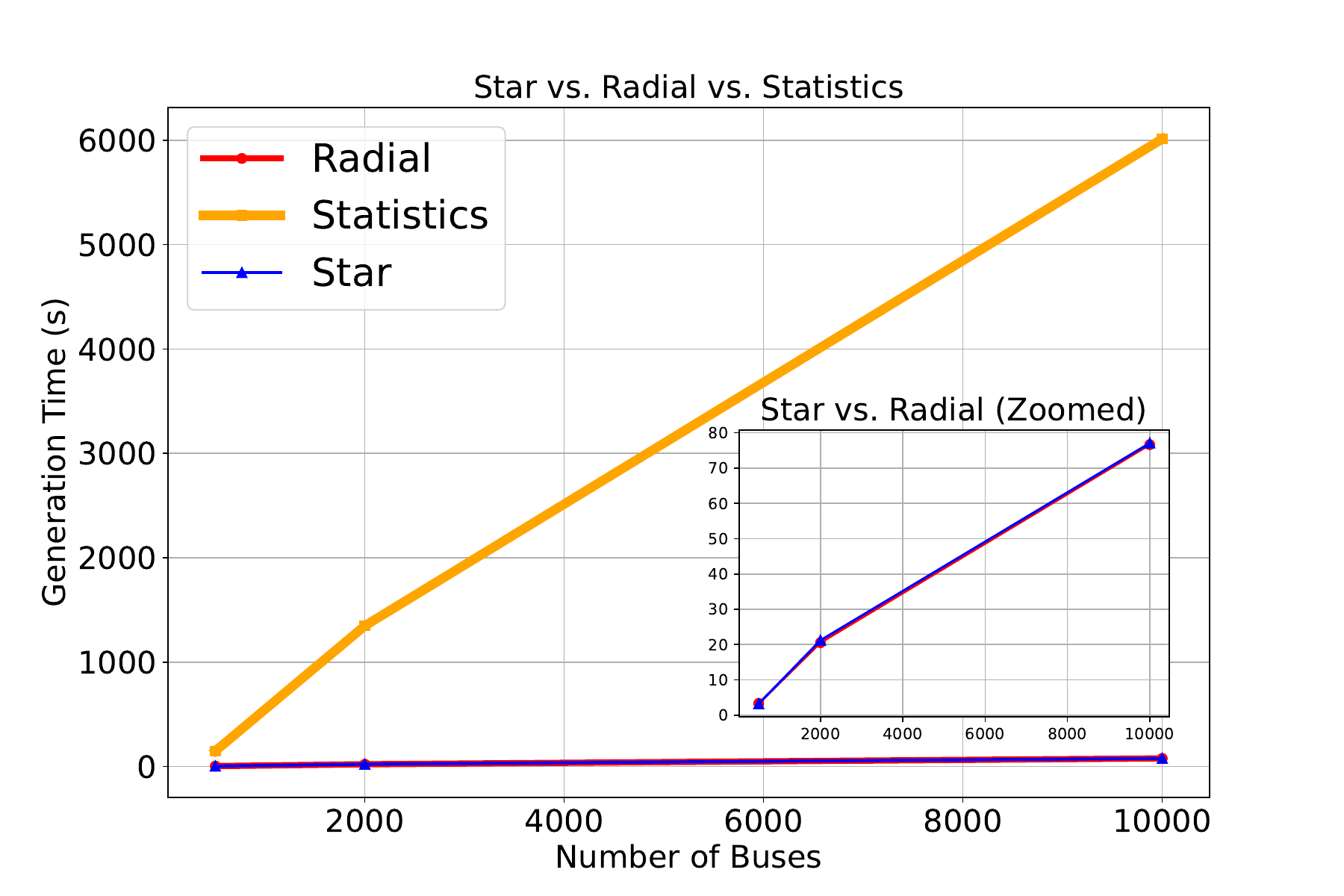}
\caption{Comparison of the number of buses and the generation time of the JSON file.}
\label{fig:BusNumVSGenTime}
\end{figure}

Metrics are chosen to compare and determine which network topologies are efficient and reliable~\cite{SandiaAnalysis}. The metrics used in this paper are relevant to researchers who study the performance, resilience, and cybersecurity of these communication network topologies, as detailed next:
\begin{enumerate}
\item \textbf{Average Path Length} ($L_{ave}$) is the average distance, measured in the number of hops, between two nodes in the network. All links are considered to have a cost of one; in other words, no weights are assigned to the links. All distances $d(x,y)$ are calculated as the shortest distance between nodes $x$ and $y$. The sum of all these shortest distances is divided by the total number of nodes $n$, as in Eq.~\ref{eq_path_length}. 
\begin{equation}
\label{eq_path_length}
L\sb{ave}=\frac{\sum_{x=1}^{n-1}\Bigr[\sum^{n}_{y=x+1}d(x,y)\Bigr]}{n}
\end{equation}

This metric is applied for all the routers in our network because distance measured in the number of hops is usually applied to count the number of routers in the path between source and destination nodes. In terms of network performance such as  delays, it is best practice to reduce the number of hops. For network reliability and cybersecurity resilience, it is desirable to have alternate paths to reach the destination. 

\item \textbf{Network Diameter} ($d$) is the greatest distance in number of hops between two nodes. This metric finds the shortest paths between all pairs of distinct nodes and selects the maximum value, as in Eq.~\ref{eq_diameter}. 

\begin{equation}
\label{eq_diameter}
d= {max[d(x,y)] | x={1,2,...,n}, y={1,2,...,n}, x\neq y}
\end{equation}

This metric is also applied only to routers. Network diameter indicates the longest path in the WAN, which is also a performance metric.

\item \textbf{Node Degree} metric quantifies the number of connections each node has, with a lower node degree being preferable because it indicates a node is connected to a smaller amount of links. In communication networks for power systems, a lower node degree means a less critical node.
On the other hand, a high node degree means potential bottlenecks or Points of Failure. 
For example, if a critical router is the target of a cyber attack, such as a Denial of Service (DoS), the communication between several substations and control center will be impacted. A DoS can impede traffic flow between nodes, by either slowing down the transmission of packets or stopping the traffic.
In order to evaluate and compare the network efficiency and resilience of different WAN topologies, the average, minimum, and maximum node degrees of routers are documented across the three use cases and topologies.
\item \textbf{Number of Links} ($m$) counts the number of links for all nodes in the generated network models. It is an essential metric because increasing the number of network links reduces path length, latency, and congestion. Yet, this expansion increases infrastructure costs and creates higher node degrees. Achieving an optimal link count requires balancing performance gains with cost and node connections.

\item \textbf{Number of Nodes} ($n$) counts all the nodes in all the Local Area Networks in NetworkX, or internal nodes, such as the total number of switches, routers, hosts, and so on. This metric gives us an idea of the size of our network, i.e., in terms of the number of network assets it has.

\item \textbf{Network Density} ($D$) refers to the number of links $m$ relative to the total number of possible links between all $n$ nodes. A denser network indicates there are more connections among nodes. This metric takes into consideration all the nodes and links the model generates. Assuming the network is an undirected graph, Eq.~\ref{eq_density} is used to calculate the network density.

\begin{equation}
\label{eq_density}
D=\frac{2m}{n(n-1)}
\end{equation}

\item \textbf{Generation Time} shows the time it takes the algorithm to generate the CPPS model. This time is measured in three pieces of the code: the JSON generation program, the NetworkX visualization program, and the file that checks the metrics. This is collected while the code runs for different test cases to verify how the generation time increases with the size of the power system test case. It is plotted in Fig. \ref{fig:BusNumVSGenTime} to help estimate the time it would take for other test cases of different sizes to run.

\item \textbf{ACL Count} quantifies the number of access control lists (ACLs) configured within the network. This count is taken for each high-level node, providing insight into the complexity of data flows and the configuration of firewalls, as depicted in Fig.~\ref{fig:protocols}.
\end{enumerate}

These metrics give us insights into the size of the network, the distance between nodes, and the performance of the network in terms of latency. These are critical for SCADA systems, for which an important requirement is response time. When a utility control center sends a command to a substation, the control center should receive a confirmation of the command or a measurement within a very short delay, in the order of milliseconds. Inside a substation, for example, the communication latency requirements are documented in the IEC~61850  protocol, as presented in~\cite{sel_sdn}. While $1 sec$ is an acceptable delay for an alarm, the delay for mission critical commands and data is between $3-10msec$~\cite{gungor}. Hence these models and metrics can help inform grid planning and cybersecurity measures

\begin{algorithm}[H]
\raggedright
\caption{Create UCC, BA Objects \& Network Topology}
\DontPrintSemicolon
\label{alg:generate_utilities}
\SetKwProg{generateutilities}{Function \emph{create\_utilities}}{}{end}
\SetKwProg{generateBA}{Function \emph{generate\_BA}}{}{end}
\generateutilities{$(Utilities\_Dict)$}{
     \For{utility $utl$ in $Utilities\_Dict$}{
         Create an instance of Utility class $utl$ \;
         Populate instance $utl$ with node objects of cyber classes\;
         [shown in Fig. \ref{fig:cyphys_classes}]\;
         $\textit{nodes} \gets$ [`cyberphysicalNodes', `cyberNodes'].\;
         $\textit{links} \gets$ [Based on the topologies in Fig. \ref{fig:star&radial}]\;
         \If{"star" in topology}{
         Add links between all substations $sub$ in $utl$.
         }
         \If{"radial" in topology}{
         Add links between gen. and trans. $sub$.\;
         Add links between all substations $sub$ in $utl$.
         }
         \If{"statistics-based" in topology}{
         $\textit{network} \gets$ Solve an optimization problem from statistics derived from a real-world utility in \cite{Stats}.\;
         Overlay \textit{network} on the power system network to match cyber nodes to power system nodes.\;
         Assign node with the highest degree as $utl$ for \textit{network}.\;
         Add links between substations based on \textit{network}.
         }
         \textbf{\textit{populate\_cyber\_information()}}\;
         Output $utl$ data into JSON files.\;
         }
}
\generateBA{$(Utilities\_Dict, num\_of\_BAs)$}{
$\textit{$Regulatories\_Dict$} \gets$ Cluster $utl$ objects into balancing
authorities using \textit{k}-means clustering with \textit{N=num\_of\_BAs}\;
    \For{regulatory $reg$ in $Regulatories\_Dict$}{
    Create an instance of Regulatory class $reg$\;
    Populate instance reg with node objects of cyber classes [shown in Fig. \ref{fig:star&radial}].\;
    \textbf{\textit{populate\_cyber\_information()}}\;
    Output $reg$ data into JSON files.\;
    }
}
\end{algorithm}

\subsection{Verification}

\begin{table*}
\centering
\caption{Comparison of metrics for all cyber-physical cases.}
\scalebox{0.85}{
 \begin{tabular}{c c c c c c c c c} 
 \hline
 \textbf{Case} & \textbf{Topology} & \textbf{$L\sb{ave}$} & \textbf{$diameter$} & \textbf{Max} & \textbf{Number} & \textbf{Number} & \textbf{Density} &  \textbf{Gen.}\\ 

 & & (hops) & (hops) & \textbf{Node} & \textbf{of} & \textbf{of} & &  \\ 

 & &  &  & \textbf{Degree} & \textbf{Links} & \textbf{Nodes} & &  \textbf{(sec)} \\
 [0.5ex] 
 \hline
\hline

SC (500)    & star   & 4.87 & 6 & 68 & 216  & 217 &  0.92\%  & 3.25\\
\hline
      & radial &  5.15 & 8 & 57 & 216  & 217 & 0.92\% & 3.35\\
\hline
      & statistic & 12.54 & 40 & 8 & 228 & 213 & 1.01\% & 147.48\\
\hline
Texas (2k) & star & 5.60 & 6 & 200 & 1290 & 1291 & 0.15\%  & 21.22\\
\hline
      & radial & 5.84 & 8 & 179 & 1316 & 1291 & 0.15\% & 20.54\\
\hline
      & statistic & 17.8 & 88 & 20 & 1368 & 1271 & 0.17\% & 1347.2\\
\hline
WECC (10k) & star & 4.32 & 5.85 & 299 & 4922 & 4942 & 0.040\% & 77.06 \\
\hline
      & radial & 4.63 & 7.8 & 275 & 5064 & 4942 & 0.041\% & 76.62 \\
\hline
      & statistic & 5.57 & 36 & 80 & 5252 & 4843 & 0.044\% & 6012.7  \\
\hline
\end{tabular}
}
\label{csv}
\end{table*}

\begin{table}
\centering
 \caption{Number of ACLs.}
 \scalebox{1.0}{
 \begin{tabular}{c c c c c} 
 \hline
 \textbf{Case} & \textbf{Reg.} & \textbf{Utility} & \textbf{Substation} & \textbf{Total ACLs}  \\  
 \hline
\hline
SC     & 4 & 224 & 624 &  852 \\
\hline
Texas & 20 & 1330 & 3750 & 5100  \\
\hline
WECC   & 80 & 5082 & 14286 & 19448 \\
\hline
\end{tabular}
}
\label{acl}
\end{table}

The outcomes for each test case across the three different topologies are detailed in Table~\ref{csv}. The average path length, denoted as $L_{ave}$, exhibits minimal variation between the star and radial topologies, typically ranging from four to six hops between routers, indicative of more direct paths. In contrast, the statistics-based topology demonstrates a considerably greater path length, averaging about 13 hops for the 500-bus case, 18 hops for the 2,000-bus case, and 6 hops for the 10,000-bus case. Additionally, the diameter $d$ of the network, which also correlates with this metric, further illustrates the disparities in topology performance. The diameter for the statistics-based topology spans 40, 88, and 36 hops for the respective bus cases, compared to only 6 to 8 hops in both the star and radial topologies. This extended path length in the statistics-based topology is a notable drawback, potentially resulting in increased latency. As described in~\cite{decentralized}, the power grid is highly interconnected and disturbances in one region can spread to other regions. This means the utility control center or a balancing authority needs to receive timely voltage and current measurements over the wide area network. A \emph{read} command sent from a UCC to a substation might experience delays due to the need to traverse more nodes, i.e., more routers, which could significantly impact the timely delivery of packets. Also, these routers may experience congestion delays. For this reason and the latency it may cause, the statistics-based topology's higher path length and network diameter are considered drawbacks of this topology, but they show more realism.

On the other hand, the maximum node degree in the statistics-based topology is significantly lower, with a peak of 8, in contrast to 68 and 57 in the star and radial topologies, respectively, for the 500-bus case. This indicates fewer critical nodes within the statistics-based topology and a lack of nodes burdened by excessive connections. As a result, the failure of a single node in the statistics-based topology would likely have a less severe impact on the communication network than a failure in either the star or radial topologies. 

\begin{figure*}[h]
\centering
\begin{minipage}[h]{0.32\linewidth}
\includegraphics[width=\linewidth]{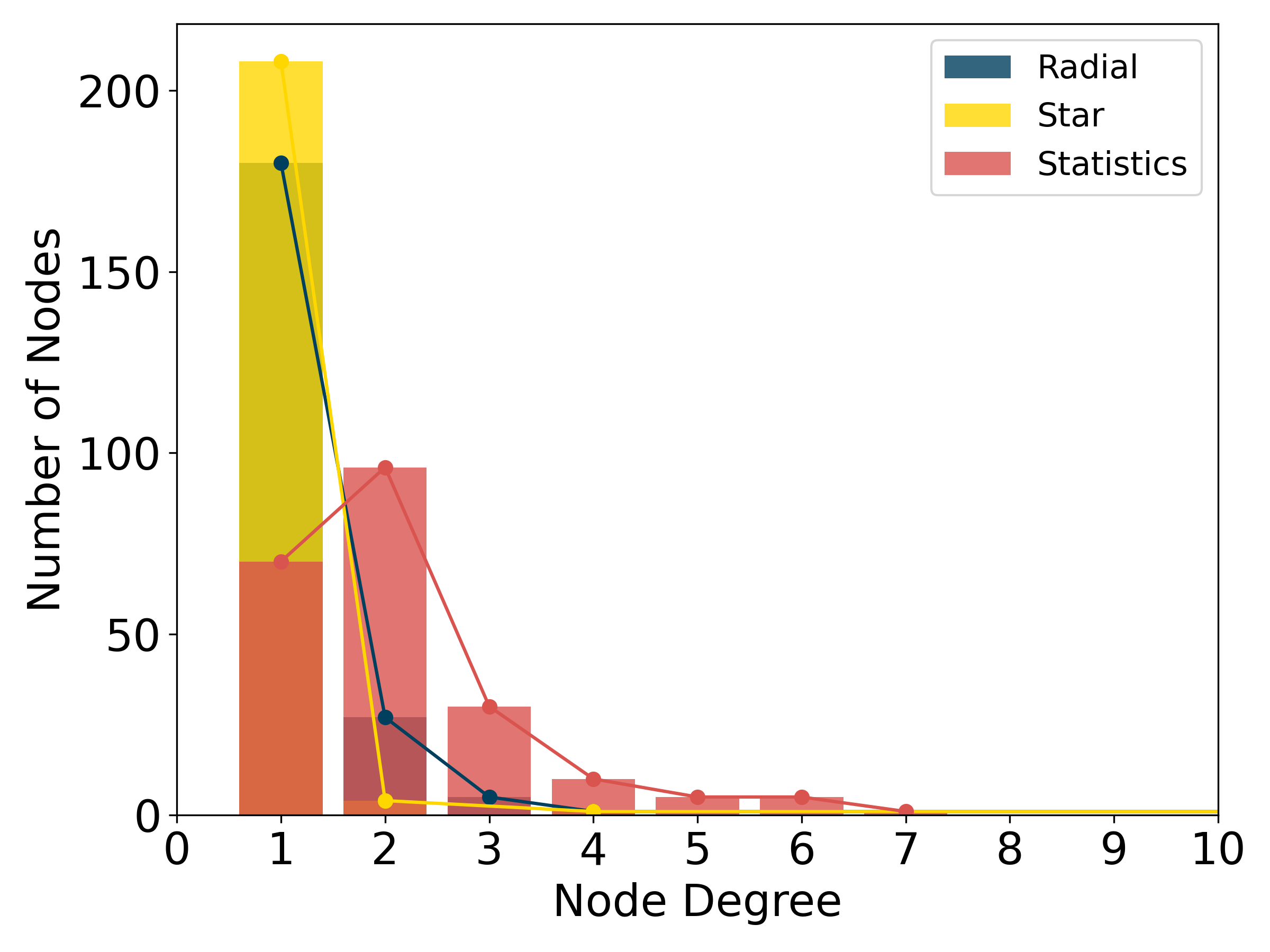}
\caption{500 bus grid}
\label{fig:ND_500}
\end{minipage}
\hfill
\begin{minipage}[h]{0.32\linewidth}
\includegraphics[width=\linewidth]{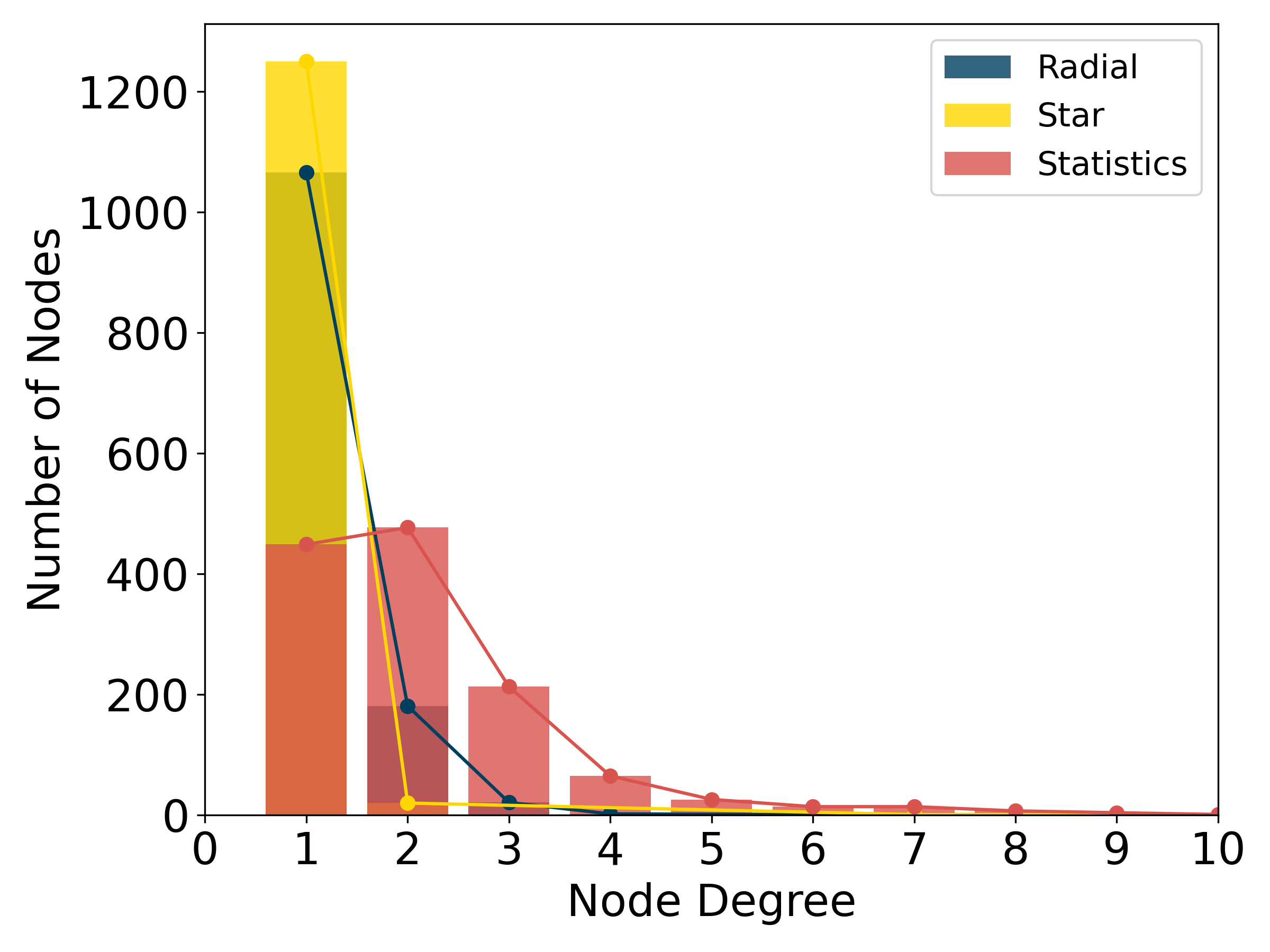}
\caption{2k bus grid}
\label{fig:ND_2k}
\end{minipage}
\hfill
\begin{minipage}[h]{0.32\linewidth}
\includegraphics[width=\linewidth]{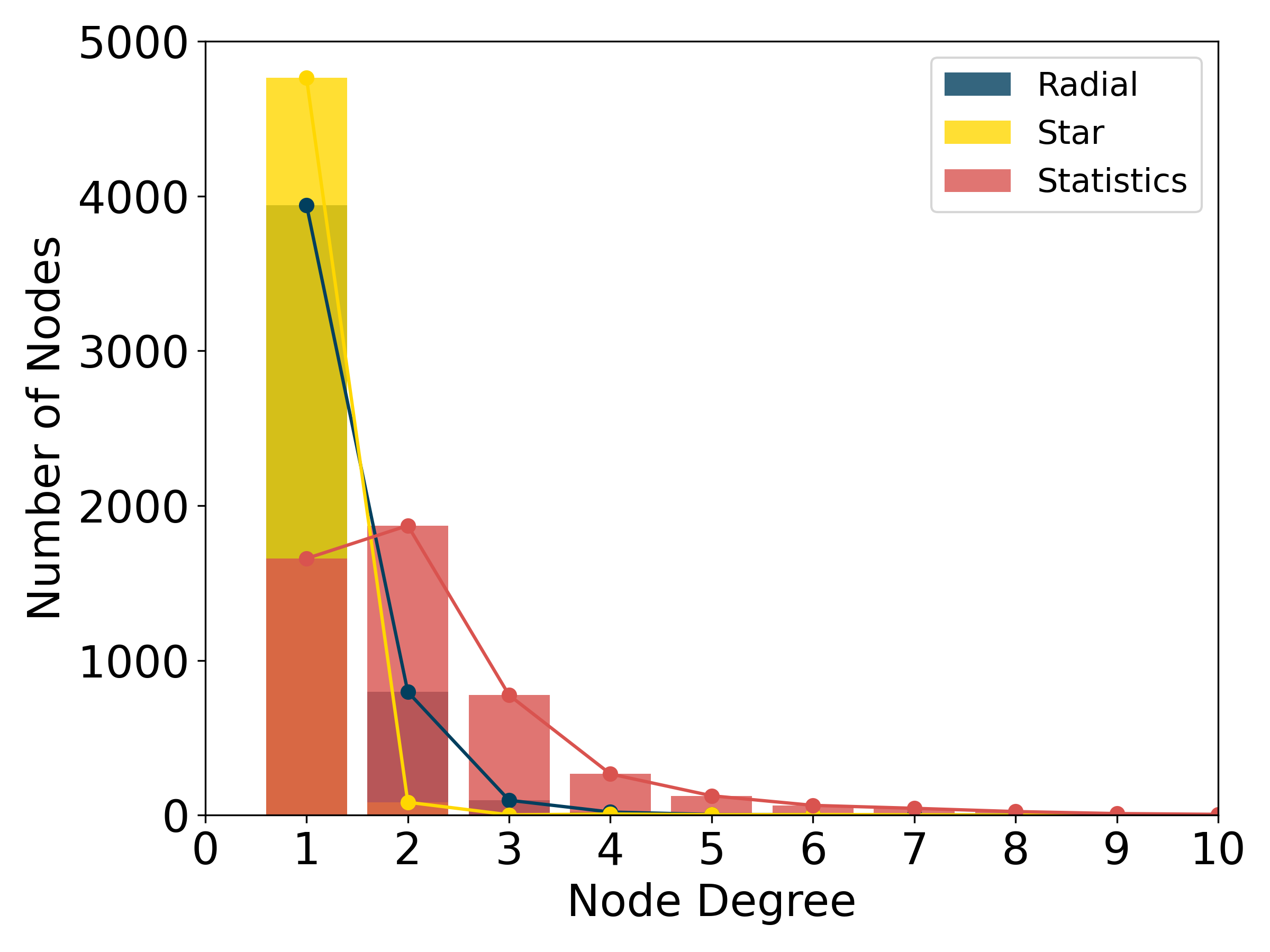}
\caption{10k bus grid}
\label{fig:ND_10k}
\end{minipage}
\end{figure*}

Node degree calculations provide insights into the connectivity of every node within the graph. Fig.~\ref{fig:ND_500}, Fig.~\ref{fig:ND_2k}, and Fig.~\ref{fig:ND_10k} illustrate a more balanced node degree distribution within the statistics-based topology, characterized by several nodes having multiple connections. Yet, none are excessively linked. In contrast, the star topology has multiple connections converge on a single node. Such bottlenecks can lead to network overload, heightened latency, and critical failure points.
The radial topology offers a slight improvement over the star topology in node degree distribution, because not all substations are directly connected to the control center. However, the radial topology still presents nodes with high degrees, such as the utility control centers.

Regarding the generation times for the cyber-physical topologies in this use case, the star topology is noted to be the quickest to produce, attributed to its simpler generation algorithm. It's important to clarify that the recorded times for running the program are taken after all distances between substations have been computed and the necessary CSV files are prepared. Initially, the program calculates the distances between substations for a given case. Once these distances are established, subsequent run times of the program are markedly reduced.

ACLs are shown in Table \ref{acl}, where the total configurations per case are determined. There is no difference between topologies because each case has the same number of BAs, UCCs, and substations per topology. This gives the user an idea of the level of complexity when setting up and configuring the cyber system for any given power system.

\subsection{Wide  Area Network}
\label{wan_networkx}
\begin{figure}[]
\centering
\begin{minipage}[h]{0.32\linewidth}
\includegraphics[width=\linewidth]{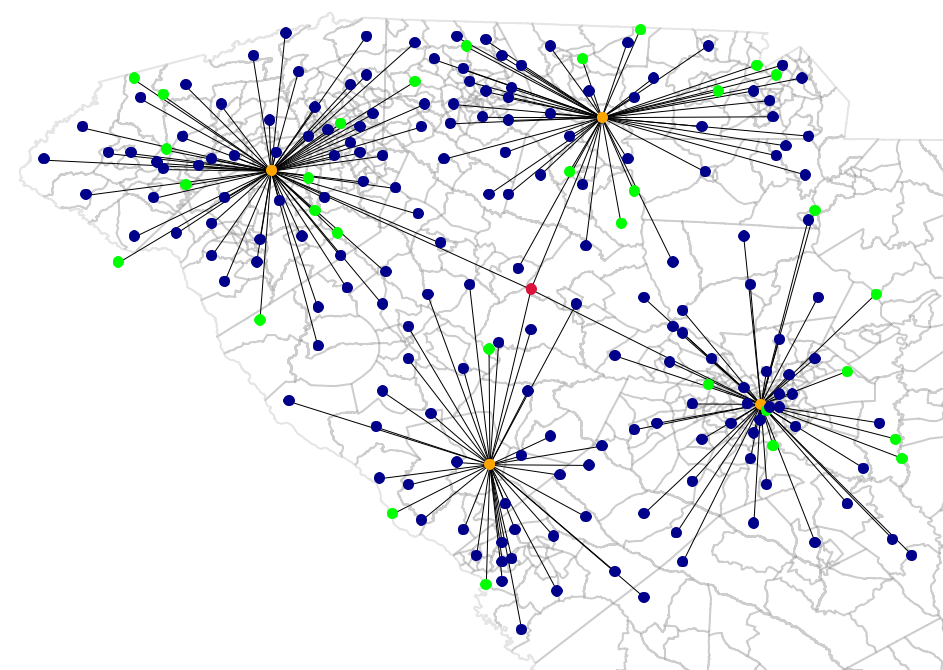}
\caption{Star Topology}
\label{fig:sc_star}
\end{minipage}
\hfill
\begin{minipage}[h]{0.32\linewidth}
\includegraphics[width=\linewidth]{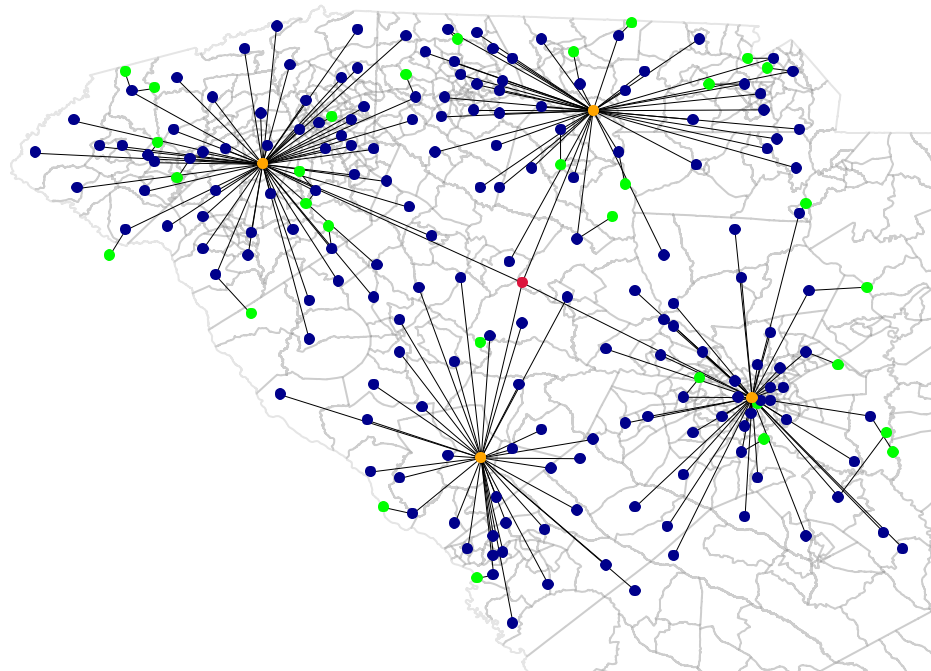}
\caption{Radial Topology}
\label{fig:sc_radial}
\end{minipage}
\hfill
\begin{minipage}[h]{0.32\linewidth}
\includegraphics[width=\linewidth]{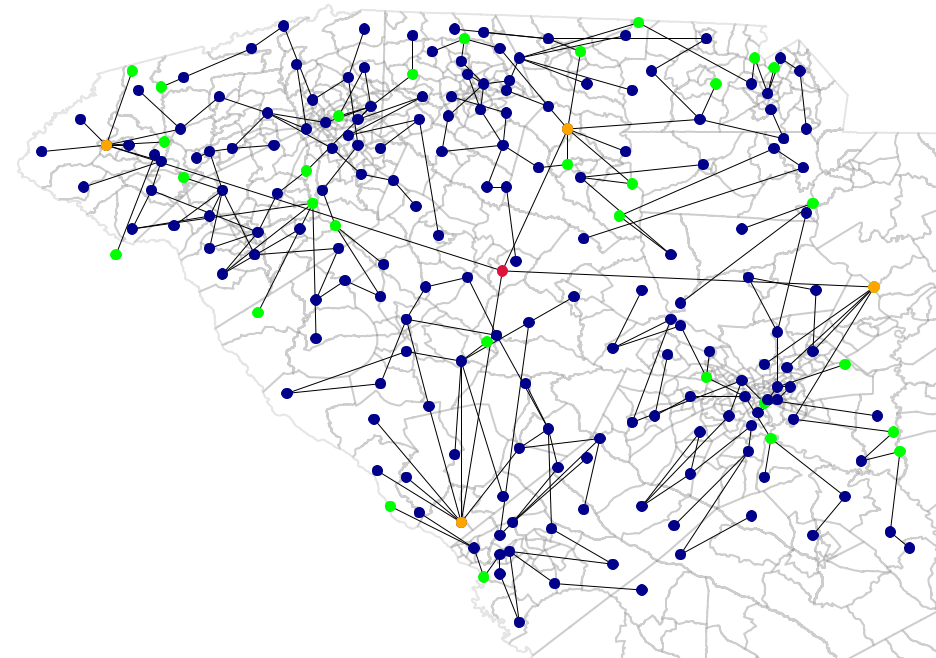}
\caption{Statistics-based Topology}
\label{fig:sc_statistical}
\end{minipage}
\caption{Cyber-physical network connection between UCCs and their respective substations for the 500-bus model on the footprint of South Carolina. The green dots are generation substations, the blue dots are transmission substations, the red dots are UCCs and the yellow dot is the BA.}
\label{fig:500networks}
\end{figure}

\begin{figure}[]
\centering
\begin{minipage}[h]{0.32\linewidth}
\includegraphics[width=\linewidth]{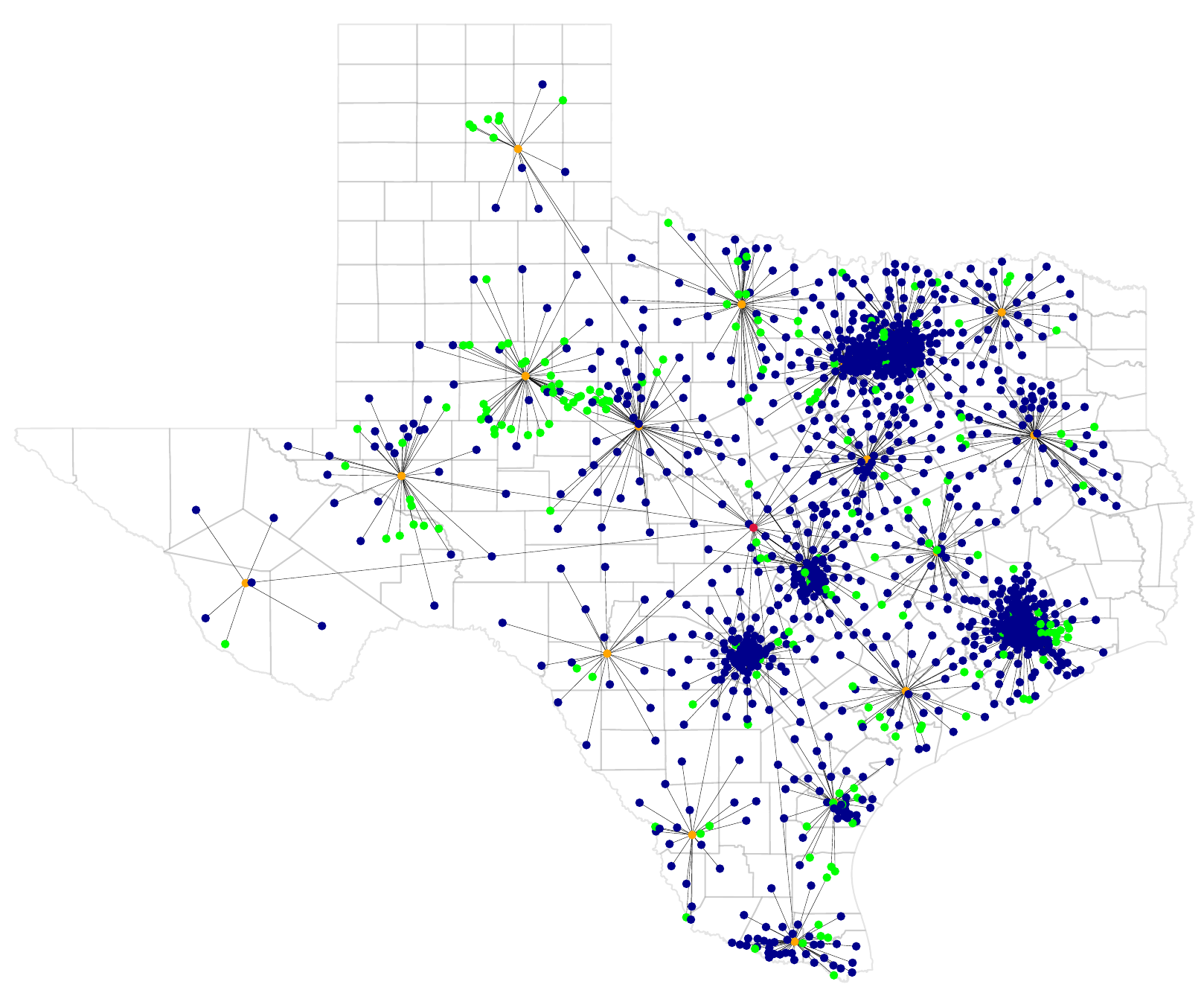}
\caption{Star Topology}
\label{fig:2k_star}
\end{minipage}
\hfill
\begin{minipage}[h]{0.32\linewidth}
\includegraphics[width=\linewidth]{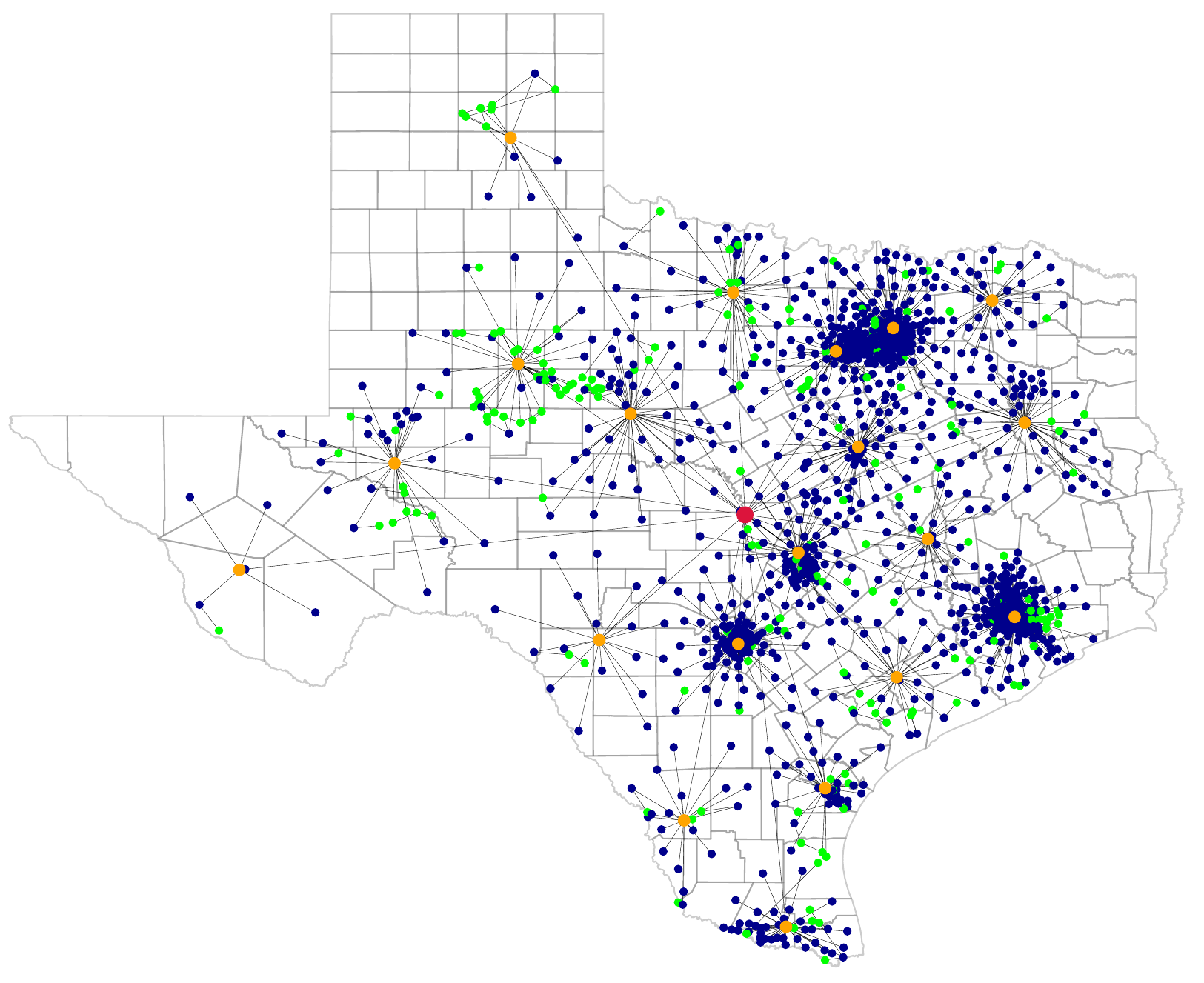}
\caption{Radial Topology}
\label{fig:2k_radial}
\end{minipage}
\hfill
\begin{minipage}[h]{0.32\linewidth}
\includegraphics[width=\linewidth]{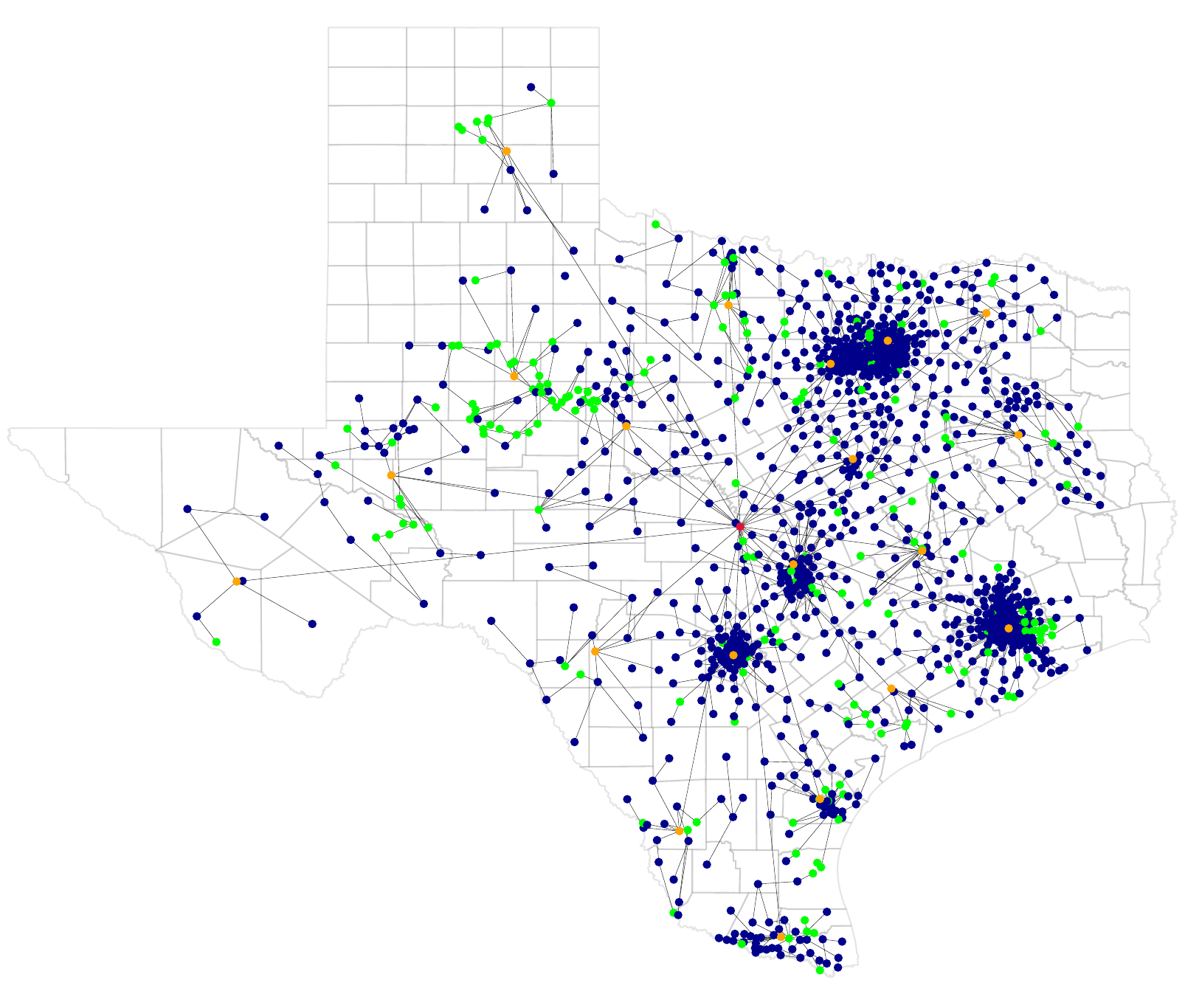}
\caption{Statistics-based Topology}
\label{fig:2k_statistical}
\end{minipage}
\caption{Cyber-physical network connection between UCCs and their respective substations for the 2k-bus model on the footprint of Texas. The green dots are generation substations, the blue dots are transmission substations, the red dots are UCCs, and the yellow dot is the BA.}
\label{fig:2knetworks}
\end{figure}

The star, radial, and statistics-based WAN topologies for the 500-bus synthetic power network are shown in Fig.~\ref{fig:sc_star}, Fig.~\ref{fig:sc_radial}, and Fig.~\ref{fig:sc_statistical}, respectively. In these figures, the yellow nodes show the location of UCCs, the blue nodes show the transmission substation locations, the green nodes show the generation substations and the single red node is the BA. Similarly, Fig.~\ref{fig:2k_star}, Fig.~\ref{fig:2k_radial}, and Fig.~\ref{fig:2k_statistical} show the same for the 2000-bus synthetic power network on the footprint of Texas. 

Of note is the significant modification of the original statistics-based algorithm in~\cite{Stats}. The original code that generates the statistics-based topology provides the optimal cyber network for any given power network. The cyber network generated from that algorithm does not contain geographical information for the cyber nodes, whereas the power network includes geographical information in its substations. To determine which cyber node belongs to which substation or its corresponding power node, the algorithm from~\cite{Stats} is modified. In this modification, the centroid of the generated cyber network is overlaid on top of the centroid of the power network. Then, the cyber network is rotated to find the Euclidean norm of the distance between each cyber node and the power node. The cyber network is rotated on top of the power network to minimize the Euclidean norm of all nodes. The overlay with the minimum value of the Euclidean norm is selected to align the cyber nodes with their corresponding substations in the power network. This approach allows adding geographic information to the cyber nodes and overlays the cyber topology onto the power system topology.

\subsection{Local Area Network}
\label{lan_networkx}
\begin{figure}[]
\centering
\begin{minipage}[h]{0.49\linewidth}
\includegraphics[width=\linewidth]{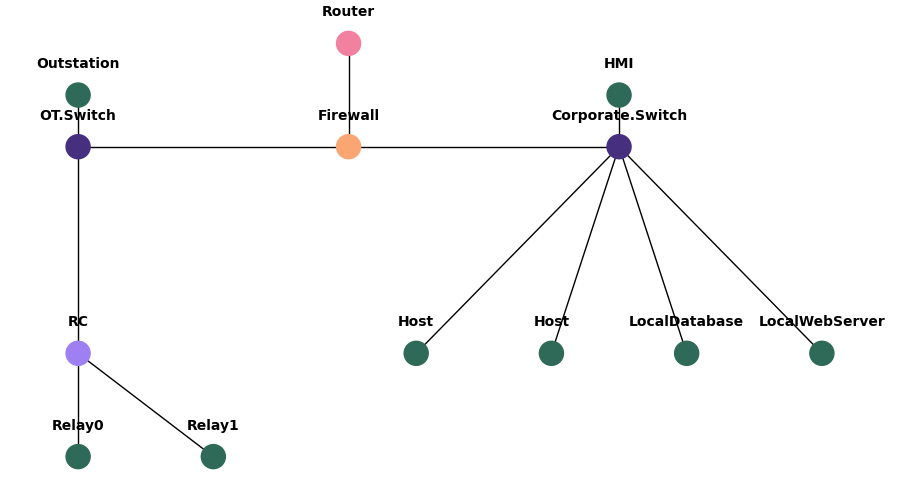}
\caption{Substation's local area networks have two separate networks, one with relays, a relay controller, and an outstation. The other with the local HMI, database, and other local hosts.}
\label{fig:lan_networkxS}
\end{minipage}
\hfill
\begin{minipage}[h]{0.49\linewidth}
\includegraphics[width=\linewidth]{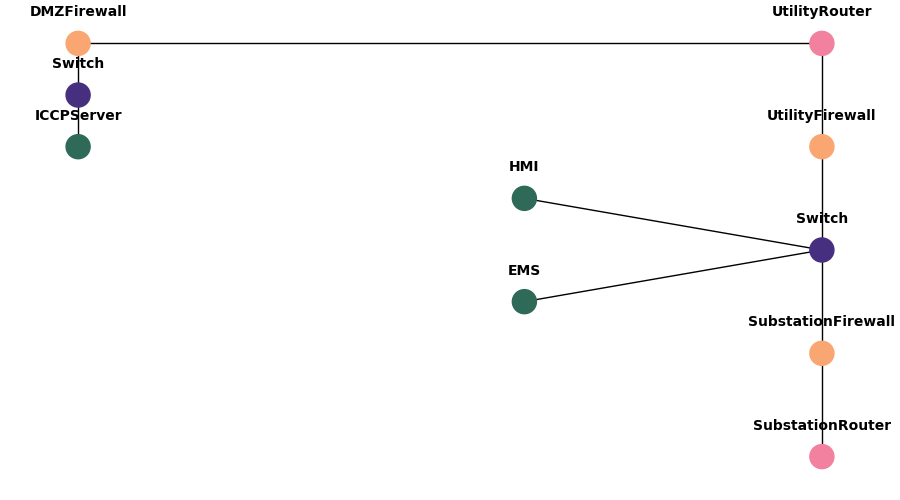}
\caption{UCCs local area network, including EMS and HMI hosts, switches, routers, firewalls, and the BA DMZ with the ICCP node. }
\label{fig:lan_networkxU}
\end{minipage}
\end{figure}

Based on the cyber models in~\cite{Texas2000}, the Local Area Network (LAN) inside one of the substations is illustrated in Fig.~\ref{fig:lan_networkxS}. This example substation has two buses, which are represented by the two relays. The nodes are color-coded by node type: routers are pink, firewalls are gold, switches are deep violet, hosts are green, and relay controllers are lavender.

The layout of nodes and links inside a substation is consistent across the different test cases with the same components used in substations and control centers, as seen in Fig.~\ref{fig:lan_networkxS} and Fig.~\ref{fig:lan_networkxU}. Each node in these figures has attributes such as utility ID, substation ID, label, and their respective VLAN.

\begin{figure}[]
\centering
\begin{minipage}[h]{0.32\linewidth}
\includegraphics[width=\columnwidth]{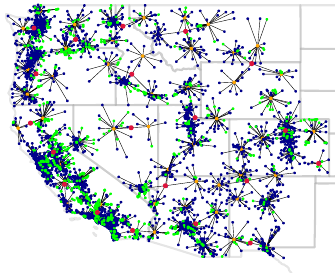}
\caption{Star Topology}
\label{fig:star_10k}
\end{minipage}
\hfill
\begin{minipage}[h]{0.32\linewidth}
\includegraphics[width=\columnwidth]{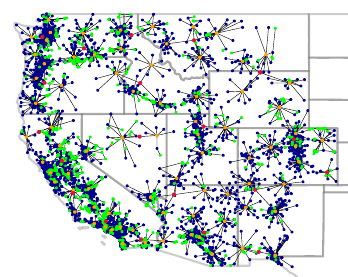}
\caption{Radial Topology}
\label{fig:star_10k}
\end{minipage}
\hfill
\begin{minipage}[h]{0.32\linewidth}
\includegraphics[width=\columnwidth]{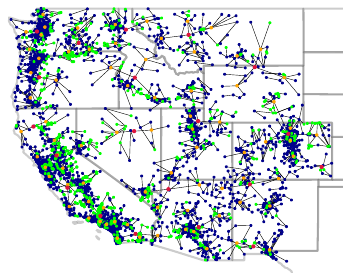}
\caption{Statistics-based Topology}
\label{fig:star_10k}
\end{minipage}
\caption{Cyber-physical network connection between UCCs and their respective substations for the 10k-bus model on the footprint of the Western United States. The green dots are generation substations, the blue dots are transmission substations, the yellow dots are UCCs, and the red dots are the BA.}
\label{fig:10knetworks}
\end{figure}

\subsection{Comparison with related work}
A related work that develops cyber-physical communication networks from a specified power grid model is the algorithm and software tool presented in~\cite{europe_topology}. This tool constructs a topology that mirrors our statistics-based approach, applying it across the power grids of several European countries. It specifically models transmission substations within the 380–400 kV high-voltage transmission network.

In Table~\ref{comparison}, there is an attempt to contrast this cyber-physical model generation tool with the approach in~\cite{europe_topology}. While there are many areas where the two approaches complement each other, the focus here is particularly on wide area network topologies and the types of data (data flows) that occur within a substation, as well as between substations, control centers, and regulatory agencies. For instance, the model generates various network topologies, including star, radial, and statistical, whereas the referenced work only supports a statistical topology. Additionally, in the radial topology, generation substations are linked to transmission substations based on the power system's branch information, as detailed in Table~\ref{csv2}.
The proposed tool creates cyber topology for a synthetic power grid with 10,000 buses spanning western U.S.. In contrast, the other covers the continental European grid with 289 nodes and some individual countries' power systems. The communication layer in the latter is split into physical and logical components, while the former integrates a logical network layer equipped with routers, firewalls, and data flows.

\begin{table}[t]
\centering
\caption{Comparing our tool with related work in~\cite{europe_topology}}
\scalebox{0.8}{
 \begin{tabular}{c c c} 
 \hline
 \textbf{Category} & \textbf{Our tool} & \textbf{The tool in~\cite{europe_topology}} \\ [0.5ex] 
 \hline
\hline
Network Topologies & star, radial, statistical  & statistical; added redundancy \\
\hline
Type of Substations & Transmission and Generation & 380–400~kV Transmission \\
\hline
Size & 500, 2k, 10k buses & Continental Europe\\
\hline
Wide Area Network & Subst., UCC, BA & Subst., Central Hub, UCC\\
\hline
Types of communication nodes & firewalls, routers, switches, hosts & routers and switches\\
\hline
Communication layer & 1 (logical) & 2 (physical and logical)\\
\hline
Data flows & DNP3, ICCP, SQL, HTTP traffic & N/A \\
\hline
\hline
\hline
\end{tabular}
}
\label{comparison}
\end{table} 

\subsubsection{Statistics-based approach}

\begin{figure*}[h]
\centering
\begin{minipage}[h]{\linewidth}
\includegraphics[width=\linewidth]{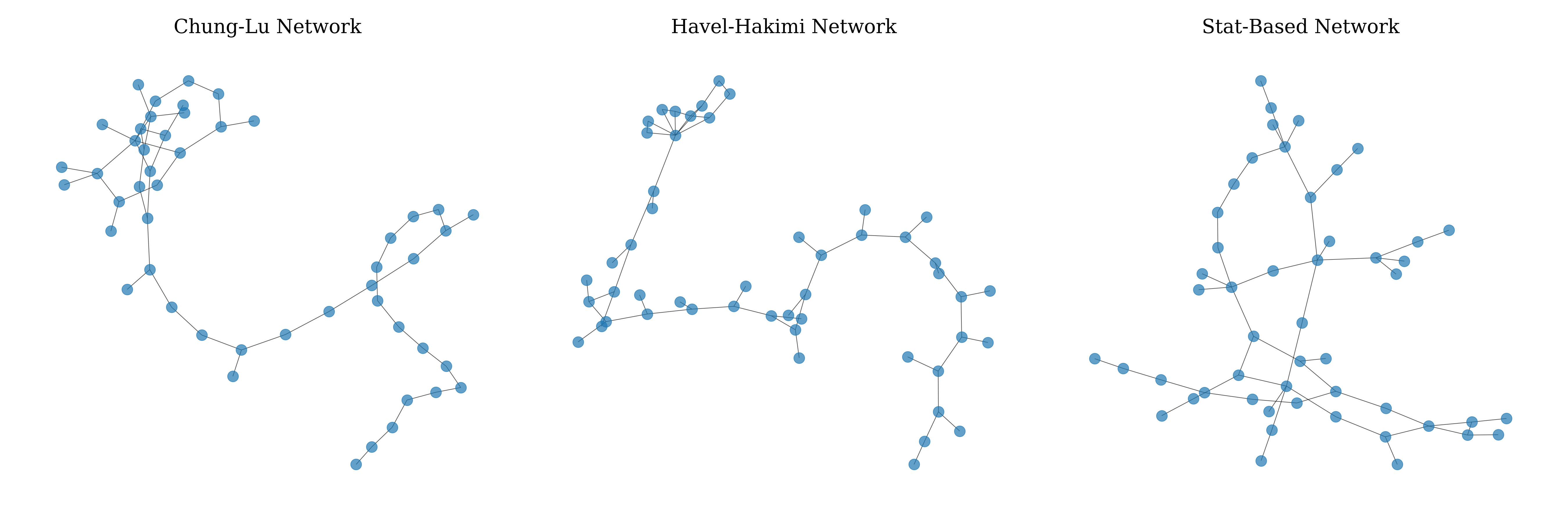}
\caption{Comparison of a 50-node network using Chung Lu,  Havel Hakimi, and statistics-based algorithm.}
\label{fig:50_comp}
\end{minipage}
\end{figure*}

\begin{figure*}[h]
\centering
\begin{minipage}[h]{0.32\linewidth}
\includegraphics[width=\linewidth]{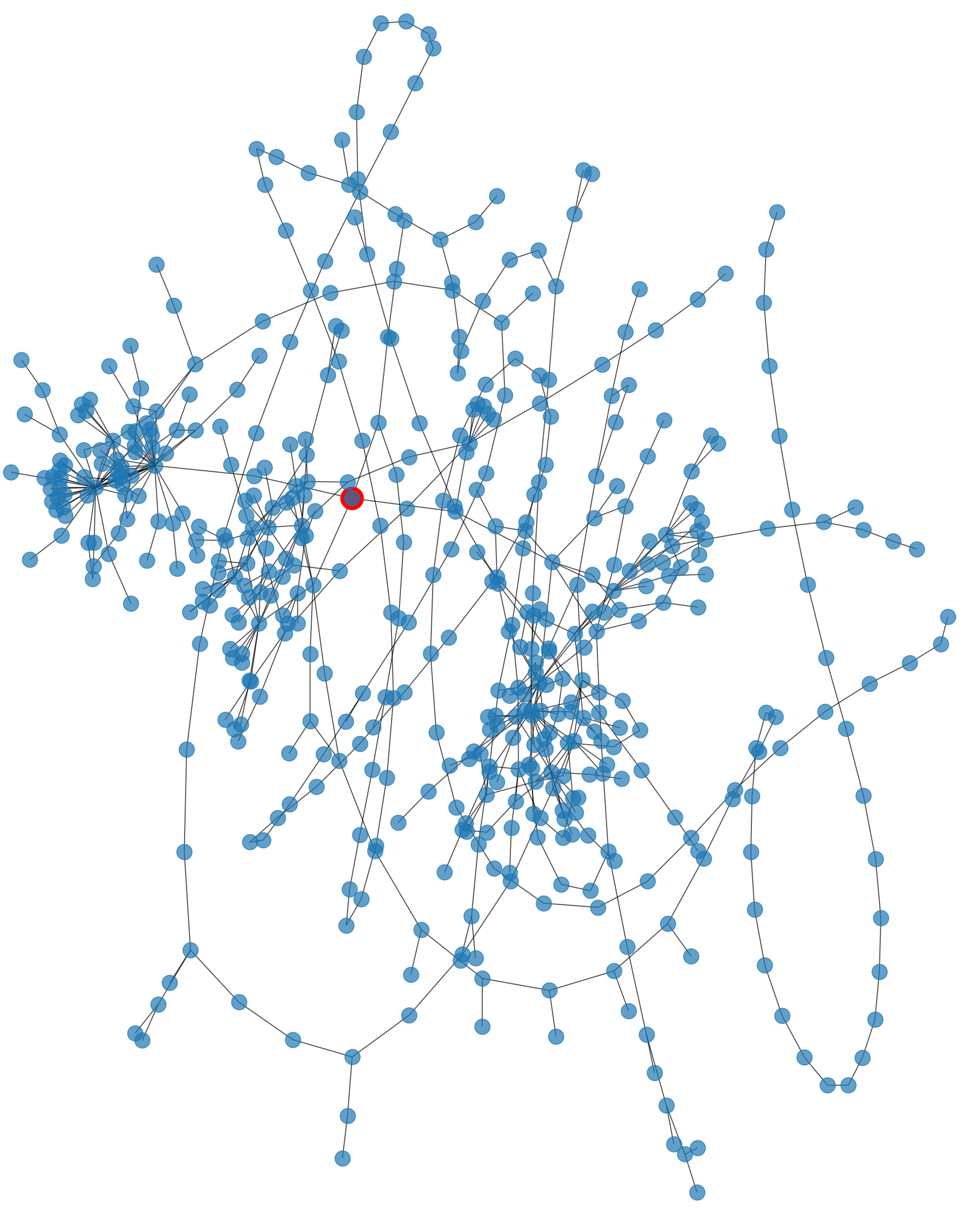}
\end{minipage}
\hfill
\begin{minipage}[h]{0.32\linewidth}
\includegraphics[width=\linewidth]{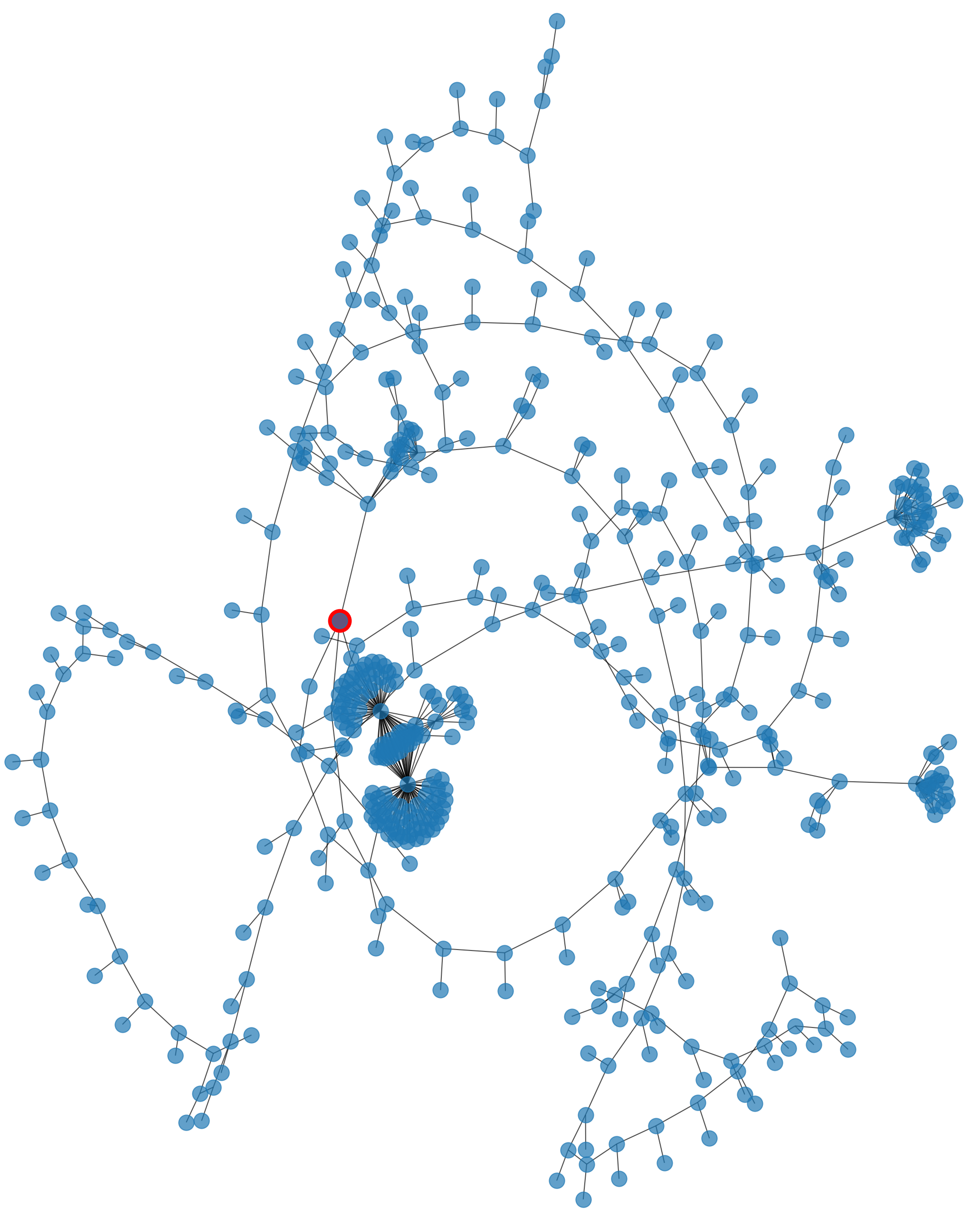}
\end{minipage}
\hfill
\begin{minipage}[h]{0.32\linewidth}
\includegraphics[width=\linewidth]{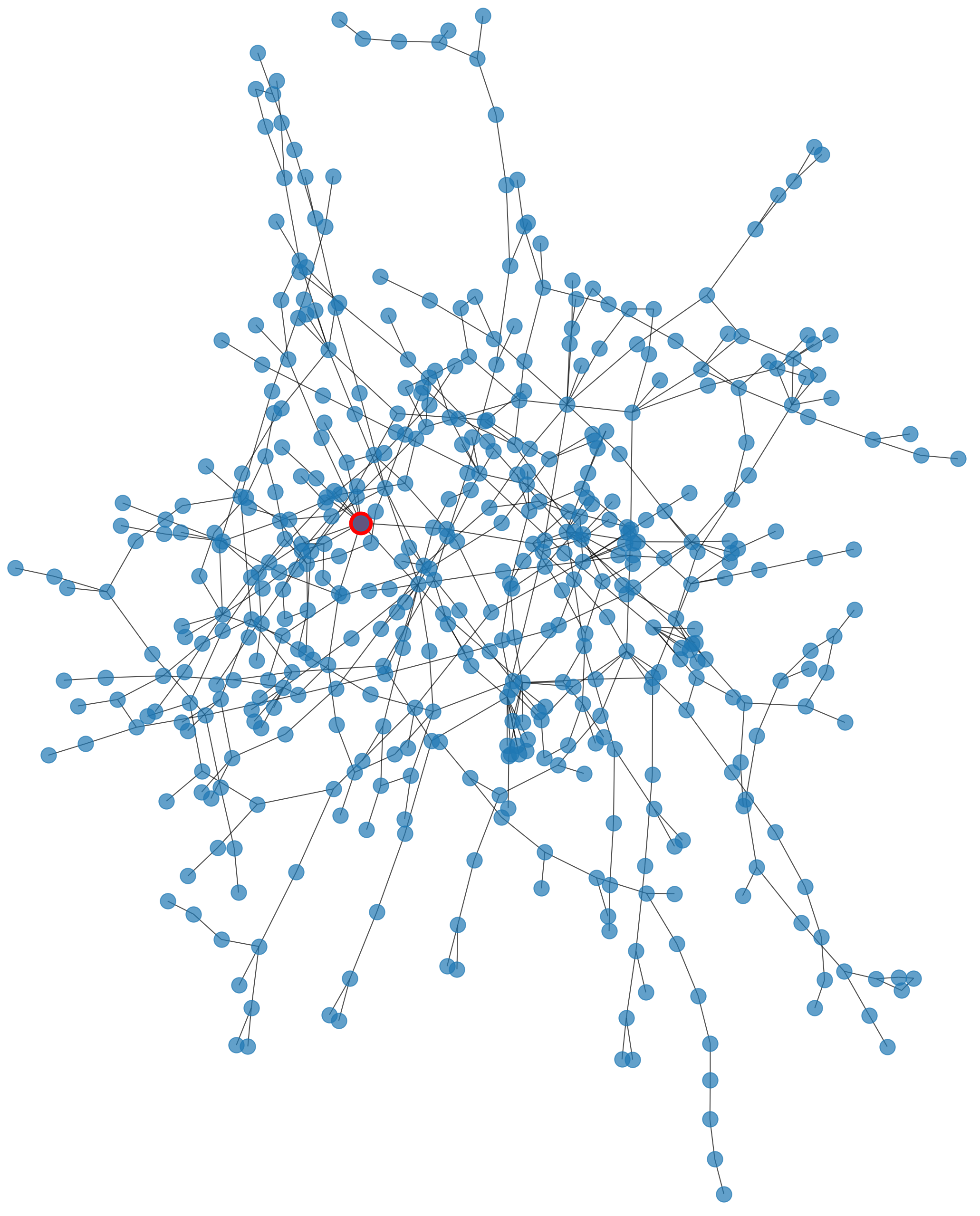}
\end{minipage}

\caption{Comparison of a 500-node network using Chung Lu(left), Havel Hakimi(center), and statistics-based algorithm(right). 
}
\label{fig:500Comparison}
\end{figure*}

The study in~\cite{europe_topology} conducted a comparative analysis of their statistics-based topology against two established graph-generation models: Havel-Hakimi and Chung-Lu~\cite{Chung}. In alignment with their methodology, these models are also employed as benchmarks for evaluating the statistics-based topology across two test cases involving 50-bus and 500-bus networks, as shown in Fig.~\ref{fig:50_comp} and~\ref{fig:500Comparison}. The graphical representations illustrate that the proposed topology achieves reduced average path lengths due to enhanced node interconnectivity. This comparison is quantified in Table \ref{tab:metrics_comparison}, which details the network metrics for both 50 and 500 node cases. Notably, the benefits of the proposed topology, characterized by decreased network delays and increased routing flexibility, become more pronounced as the network size expands. These results underscore the superiority of the proposed statistics-based topology in representing cyber-networks.

The Chung-Lu algorithm generates random graphs that align with a given expected degree sequence, making it particularly relevant for power grids where nodes—such as substations, generators, or loads—exhibit specific connectivity patterns. These patterns must be precisely replicated in synthetic networks to accurately simulate real-world behaviors. The Chung-Lu model simplifies the creation of graphs by connecting nodes randomly, with the connection probability between any two nodes being proportional to the product of their expected degrees. In contrast, the Havel-Hakimi algorithm is employed to produce graphs that exactly match a specified degree sequence. This accuracy is crucial for power grids, where the exact number of connections per node is essential for maintaining system realism.

However, these algorithms do not account for clustering, community structure, or other higher-order network properties typical of real-world networks, such as those found in power grid utilities. Graphs generated with a random structure may be adequate for analyzing smaller networks but do not accurately reflect the geographical or functional constraints of actual large-scale power systems. Consequently, our proposed statistics-based topology introduces real-world utility-based metrics to capture the complexities of power grid networks accurately.

\begin{table*}
\centering
\caption{Comparison of proposed statistics-based topology with Havel-Hakimi and Chung-Lu models.} 
\scalebox{0.8}{
 \begin{tabular}{c c c c c } 
 \hline
 \textbf{Case} & \textbf{Topology} & \textbf{$L\sb{ave}$} & \textbf{$diameter$} & \textbf{Max}  \\

 & & (hops) & (hops) & \textbf{Node}   \\ 

 & &  &  & \textbf{Degree}  \\  
 [0.5ex] 
 \hline
\hline

50-node    & Statistics-Based & 8.4 & 24 & 5    \\
\hline
      & Havel-Hakimi & 8.62 & 22 & 10 \\
\hline
      & Chung-Lu & 7.72 & 22 & 10 \\
\hline
SC (500)  & Statistics-Based & 12.54 & 40 & 8 \\
\hline
      & Havel-Hakimi & 27.6 & 85 & 69 \\
\hline
      & Chung-Lu & 27.04 & 86 & 43  \\
\hline
\end{tabular}}

\label{tab:metrics_comparison}
\end{table*}

\section{Conclusions}
\label{Section:Conclusions}
Realistic high-fidelity cyber-physical system models are important for understanding and improving the security of large-scale electric power systems. Machine readable and simulation/emulation-capable complete models can greatly enhance capabilities for defense of critical infrastructures, by improving coordinated engineering assessments of the system leading to heightened situational awareness and decision-making at all stages of an event life cycle. This is crucial for promoting improved techniques and their validation for planning, detection, and response, to assess, avoid, and mitigate the potential impacts of adversarial actions on the grid.

An algorithm for generating a comprehensive cyber-physical model was developed in \name, offering a versatile approach that can accommodate any power system model and transform it into a cyber-physical network, presenting the results in the form of a JSON file. The algorithm's applicability and effectiveness are demonstrated through its implementation in three fictitious power systems: a 500-bus test case situated in South Carolina, a 2000-bus test case in Texas, and a 10,000-bus test case on the Western United States. Each test case was modeled using three different network topologies such as star, radial, and statistics-based topology. One notable feature of this work is its generality, making it suitable for adaptation to a wide range of power system networks.

To model the data flows, four different communication protocols are used in this current model as it can be easily altered for future use. We aim to improve the algorithm's versatility by allowing users to specify communication protocols as input. This enhancement recognizes that communication protocols can vary based on the utility's location and type, making it a valuable user-configurable feature. Another configurable feature should be the relay connections within a substation, partially determined by the substation topology. 

In future work, we plan to improve how utility control center locations are chosen and which substations are connected to them. When using the radial topology, we noticed that some generation substations are connected to multiple transmission substations. We can use this information to group all these substations under the same utility control center. 

Although we did not evaluate SAM-GT on resource-constrained devices, it can generate the JSON files using a regular general-purpose personal computer. One practical deployment of the cyber-physical models is to integrate SAM-GT's code to an energy management system (EMS) host, such as the next-generation cyber-physical EMS proposed in~\cite{cypresjournal2023}. This EMS currently supports one test case with one topology: the Texas-2000 bus model with star topology. Thus, integrating SAM-GT in this cyber-physical EMS will allow researchers to model and study multiple cyber-physical test cases with the option to select topology based on research needs.

\section*{Acknowledgment}
The authors would like to acknowledge the  US Department of Energy under award DE-CR0000018.

\bibliographystyle{ACM-Reference-Format}
\bibliography{GenModelBib}
\end{document}